\documentclass{article}[11pt]
\usepackage{amsmath}
\usepackage{amsfonts}
\usepackage{amssymb}
\usepackage{graphicx}
\usepackage{epstopdf}
\usepackage{cite}
\usepackage[usenames]{color}
\usepackage{graphicx}
\usepackage{epstopdf}
\pdfoutput=1
\providecommand{\U}[1]{\protect\rule{.1in}{.1in}}
\newcommand{\ba}{\begin{array}}
\newcommand{\ea}{\end{array}}
\newcommand{\Dsl}[1] { \setbox0=\hbox{$#1$}     
\dimen0=\wd0   \setbox1=\hbox{/} \dimen1=\wd1  \ifdim\dimen0>\dimen1        
 \rlap{\hbox to \dimen0{\hfil/\hfil}}  #1 \else \rlap{\hbox to \dimen1{\hfil$#1$\hfil}}  /  \fi  }
\newcommand{\bea}{\begin{eqnarray}}
\newcommand{\eea}{\end{eqnarray}}

\newcommand{\ns}{\Dsl{n}}
\newcommand {\nbs}{\Dsl{\bar n}}

\newcommand{\ps}{\Dsl{p}}
\newcommand{\ks}{\Dsl{k}}
\newcommand{\qs}{\Dsl{q}}

\input{tcilatex}

\begin{document}

\title{ {\Large
 Radiative decays $\chi_{cJ}\rightarrow V\gamma $ within the  QCD factorisation framework
}}
\author{ N. Kivel \thanks{%
On leave of absence from St.~Petersburg Nuclear Physics Institute, 188350,
Gatchina, Russia} \ and M. Vanderhaeghen \\
%EndAName
\textit{Helmholtz Institut Mainz, Johannes Gutenberg-Universit\"at, D-55099
Mainz, Germany} \\
\textit{Institut f\"ur Kernphysik, Johannes Gutenberg-Universit\"at, D-55099
Mainz, Germany } }
\date{}
\maketitle

\begin{abstract}
We present a discussion of  radiative decays $\chi_{cJ}\rightarrow\gamma \rho (\omega, \phi)$. The decay amplitudes  are computed within the QCD factorisation
framework.  NRQCD has been used  in  the heavy meson sector  as well as a collinear expansion  in order to describe the overlap with  light  mesons in the final state. 
The  colour-singlet  contributions to all  helicity amplitudes  have been computed using the light-cone distribution amplitudes of twist-2 and twist-3. 
All obtained expressions are  well defined at least in the leading-order approximation. The  colour-octet contributions have also been studied  in the Coulomb limit  in order to exhibit  their scaling behaviour.  In order to understand the relevance of the  different  contributions we perform numerical estimates using  the colour-singlet contributions.  
For the $\chi_{c1}\rightarrow \gamma V_\perp$  decays, to vector mesons with transverse polarisation,  we find that  the colour-singlet contribution potentially allows  for a reliable  description. 
On the other hand, for the  $\chi_{c1}\rightarrow \gamma V_\Vert$ decays, to vector mesons with longitudinal polarisation, our findings  indicate that the colour-octet mechanism may be important for a good description.   We expect  that  more accurate  measurements of the decay $\chi_{c2}\rightarrow\gamma V_{\perp,\Vert} $  can help to better  understand the mechanism of radiative decays.        
\end{abstract}

\noindent

\begin{center}

%\vspace*{1cm}

%HIM-2013-XX \bigskip

\vspace*{1cm}

% \textbf{DRAFT, \today}

\end{center}

\newpage

\section{Introduction}
\label{int}

The  radiative decays of  $P$-wave  charmonia $\chi_{cJ}$ have  been measured  by different experimental collaborations:  CLEO \cite{Bennett:2008aj}  and BESIII \cite{Ablikim:2011kv}.   
Theoretical estimates have been given  in Refs.\cite{Gao:2006bc, Gao:2007fv}.   In these  papers, the authors use the pQCD 
formalism in combination with   specific models for the light meson wave functions,  and  the constituent quark masses have been  used  as infrared
regulator in these calculations.   The obtained estimates for the  $\chi_{c1}\rightarrow \rho(\omega, \phi)\gamma$  decay rates are few times smaller than 
the measured ones. To resolve this discrepancy the authors of  Ref.\cite{Chen:2010re}  used   a phenomenological model with  intermediate $D-$meson interactions. 
 
\begin{figure}[h]
\centering
\includegraphics[width=5.0 in]
{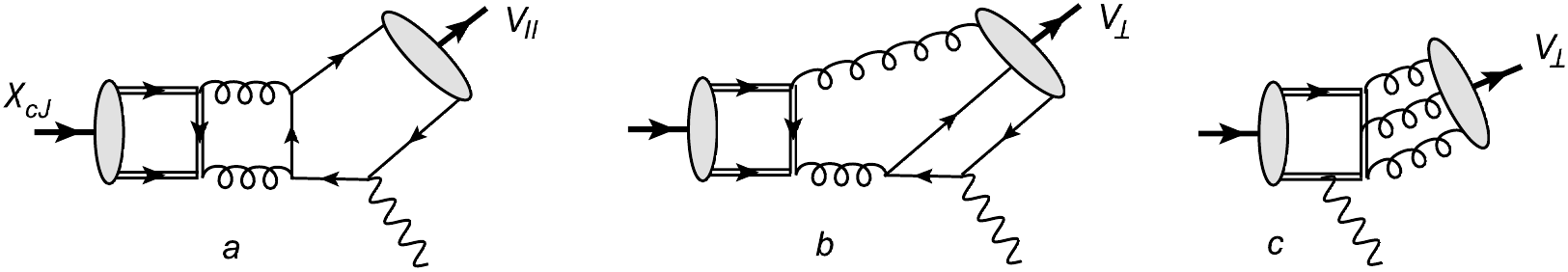}
\caption{ The various contributions to the decay amplitudes: $(a)$ the leading-order contribution for the longitudinal amplitude with $V_\Vert$; $(b)$ and $(c)$  two different contributions for the transverse amplitude with $V_\perp$.  }
\label{2g-hard}
\end{figure}

 In Table~\ref{dataXc1}   we collect the current branching fraction measurements for the decays which have been measured.  From this table it is seen that  the largest  branchings for all channels  are provided by decay  $\chi_{c1}\rightarrow V\gamma$.  Moreover, in this case the decay rate is dominated by  the longitudinal meson  ($V_\Vert$) in the  final state.  
\begin{table}[th]
\centering
\begin{tabular}{|c|c|c|c|c|c|}
\hline
 & $ \chi _{c1}\rightarrow V \gamma$  & $\chi _{c1}\rightarrow V_{\Vert }\gamma $ 
 & $\chi _{c1}\rightarrow V_{\bot }\gamma $ & $\chi _{c0}\rightarrow V\gamma  $ &  $\chi _{c2}\rightarrow V\gamma$ \\ \hline
$\gamma \rho $ & $220\pm 18$  & $184.8\pm 15.7$ & $35.2\pm 7.4$ & $ <9 $ & $<20$
 \\ \hline
$\gamma \omega $ & $69\pm 8$ & $51.8\pm 8.9$ & $17.3\pm 6.5$ &$<8$  & $<6 $
\\ \hline
$\gamma \phi $ & $25\pm 5$  & $17.7\pm 4.9$ & $7.3\pm 3.6$ &$<6 $ & $<8$
\\ \hline
\end{tabular}%
\caption{ The branching fractions  $\chi _{cJ}\rightarrow V\gamma $  in units of $10^{-6}$. The total fractions are  taken from the Review of Particle Physics \cite{Olive:2016xmw}. 
The original experimental results  can be found in \cite{Bennett:2008aj, Ablikim:2011kv}. 
In order to obtain the different contributions for the $\chi_{c1}$ decays we used the ratios $f_{\bot }^{V}={\Gamma \lbrack \chi _{c1}\rightarrow V_{\bot }\gamma ]}/
{\Gamma \lbrack \chi _{c1}\rightarrow V\gamma ]}$ from Ref.\cite{Ablikim:2011kv}.
}
\label{dataXc1}
\ \ 
\end{table}
The data presented in Refs.\cite{Bennett:2008aj,Ablikim:2011kv} also  allow one  to study the contributions of the different helicity amplitudes which can
 provide  additional interesting information about the  underlying  mechanism of  quark-gluon interactions. This point  has not yet been considered   to  full extent in the literature.  For instance,  within the systematic  QCD factorisation framework  the leading-order contribution with a longitudinal outgoing vector meson  is given by the  diagram as in Fig.\ref{2g-hard}$(a)$  but  for a transversely polarised  meson ($V_\perp$) one has to consider the matrix element with the three particle wave functions  as in Fig.\ref{2g-hard}$(b)$ and Fig.\ref{2g-hard}$(c)$. The first diagram is of order $\alpha_s^2$, the second is  of order  $\alpha_s$  but suppressed by a factor $\Lambda_{QCD}/m_c$ because of  subleading twist-3 collinear operators describing the overlap with the light outgoing meson.    

 A description of the reactions with charmonium states  can be challenging because of possible large contributions of the colour-octet  operators; see, e.g.,  review \cite{Brambilla:2004wf} and references therein.  This mechanism  could  be especially important  for the description of exclusive $P$-wave hadronic decays and  has been studied within a phenomenological framework in 
 Refs.\cite{Bolz:1996wh, Bolz:1997ez} and  in the Coulomb limit  in Ref.\cite{Beneke:2008pi}.
 The contributions with the colour-octet operators  can also play  an important role in the description of the radiative decays of  $\chi_{cJ}$.  A hint about this  can be seen from the following  observation. 
 
The various contributions to the decay amplitudes can be associated with the following two matrix elements: either the photon is emitted from the light quark  or  from the heavy quark
\bea
 \langle V(p)|  J^\mu_{em} |\chi_{cJ}\rangle =  \langle V(p)|  \sum_{u,d,s }e_q\  \bar q \gamma^\mu q |\chi_{cJ}\rangle +  \langle V(p)| e_c\  \bar c \gamma^\mu c |\chi_{cJ}\rangle
\label{Jem}
\eea
If the  dominant  contribution to the decay amplitudes is provided  by the light quark component of the electromagnetic current in Eq.(\ref{Jem}) then using  $SU(2)$  in the light quark sector one  can  establish  the relations between the  branching fractions for different vector mesons in the final state. For instance, one easily finds  the relation  
${Br\left[ \chi _{c1}\rightarrow \omega \gamma \right] }/{Br\left[ \chi _{c1}\rightarrow \rho\gamma \right] }\simeq \frac{1}{9}$ .  This  prediction  can be easily understood  if one assumes that  the decay amplitude  is dominated by the hard two-gluon intermediate state as in Fig.~\ref{2g-hard}$a$.  The same arguments   must  also be valid for the colour-octet  contribution which originates from  the light-quark matrix element in Eq.(\ref{Jem}).   However,  as one can see from the data in Table \ref{dataXc1},  the given ratio  is about a factor 3 larger than expected.  This allows  one to  assume that  the heavy quark matrix element  in Eq.(\ref{Jem})  also gives contributions with a sizable numerical impact.  However in the factorisation framework the corresponding contribution with  longitudinal mesons $V$  can  only  be   associated  with the  colour-octet mechanism.

 In order to better  understand the relevance  of the various contributions,  in this work  we compute  the  helicity amplitudes within the standard
 QCD factorisation framework and study the possible colour-octet contributions in the Coulomb limit.  
 The factorisation   approach  is closely related  with the  effective theories  inspired by QCD: NRQCD \cite{Lepage:1992tx, Bodwin:1994jh, Brambilla:2004jw}  for the heavy quark sector and 
the  soft-collinear effective theory (SCET) \cite{Bauer:2000ew, Bauer2000, Bauer:2001ct, Bauer2001, Beneke:2002ph, Beneke:2002ni}  for the collinear sector associated with the  light  mesons in the final state.  The important advantage of this 
scheme is that it allows one to perform systematic expansions with respect to the small relative velocity $v$ and $\lambda^2 \sim \Lambda_{QCD}/m_Q$.  Hence
one can classify various operators  with respect to these parameters   and this can guide us on the relevance of the different contributions.

The outline of this paper is as follows. In Sec.~\ref{hard} we provide the  basic notations and  description of the decay amplitudes and decay width.  Then we  compute  the contributions of the colour-singlet operators for  different polarisations of the outgoing light meson. We show that the leading-order contributions are factorizable
and establish their scaling behaviour.   This  behaviour can be qualitatively compared with  the behaviour of the  colour-octet contributions in the Coulomb limit.
 This  allows us to make a qualitative conclusion about the importance  of the colour-octet  matrix elements.  
In Sec.~\ref{phen} we perform numerical estimates of the branching fractions using only the colour-singlet  approximations for the decay amplitudes.  This analysis allows us 
to make some more realistic  conclusions about possible  mechanism of the radiative decays. The summary of our results is presented  in Sec.~\ref{conc}. In  Appendixes~A and B  we provide  a  description of the different vector meson distribution amplitudes and present  technical details on the  estimates of the colour-octet  matrix elements in the Coulomb limit.

\section{Colour-singlet contributions in the factorisation framework}
\label{hard}

\subsection{Kinematics and amplitudes. }

The amplitude of decay $\chi
_{cJ}(P)\rightarrow \gamma (q)+V(p)$ is defined as%
\begin{equation}
\left\langle \gamma (q)V(p);\text{out}\right\vert \left. \text{in};\chi
_{cJ}(P)\right\rangle =i(2\pi )^{4}\delta (P-p-q)\ \mathcal{A}_{\chi
_{cJ}\rightarrow V\gamma },
\end{equation}%
where%
\begin{equation}
\ \mathcal{A}_{\chi _{cJ}\rightarrow V\gamma }=\left\langle V(p)\right\vert
\epsilon _{\gamma }^{\ast }\cdot J_{em}(0)\left\vert \chi
_{cJ}(P)\right\rangle ,  \label{Ame}
\end{equation}
with electromagnetic current 
\begin{equation}
J_{em}^{\mu }(0)=\sum_{q=u,d,s,c}ee_{q}\bar{q}(0)\gamma ^{\mu }q(0),
\end{equation}
here $\epsilon _{\gamma }^{\ast }$ is the photon polarisation vector. In the
following we use the frame where the heavy meson is at rest and the $z$-axis is
chosen along the momenta of the outgoing particles
\begin{equation}
P=M_{J}(1,\vec{0})=M_{J}\omega ,  \label{def:w}
\end{equation}%
where $M_{J}$ is the mass of the heavy quarkonia and $\omega $ denotes its
four-velocity. The outgoing momenta read%
\begin{equation}
q=E_{\gamma }(1,0,0,1)=2E_{\gamma }\frac{\bar{n}}{2},~\ p=(\sqrt{m_{V
}^{2}+E_{\gamma }^{2}},0,0,-E_{\gamma }),~~E_{\gamma }=\frac{%
M_{J}^{2}-m_{V}^{2}}{2M_{J}}.
\end{equation}%
Using that the heavy quark mass is quite large $m\gg \Lambda _{QCD}$ one
finds%
\begin{equation}
~q\simeq m(1,0,0,1)=2m\frac{\bar{n}}{2},~\ \ p\simeq m(1,0,0,-1)=2m\frac{n}{2%
},
\end{equation}%
where we introduced auxiliary light-cone vectors $n$ and $\bar{n}$ with $(n%
\bar{n})=2$. \ The four-velocity in Eq.(\ref{def:w}) reads%
\begin{equation}
\omega =\frac{1}{2}\left( n+\bar{n}\right) ,~\ \omega ^{2}=1.
\end{equation}%
Any four-vector $F^{\mu }$ can be expanded as 
\begin{equation}
F^{\mu }=\left( F\cdot n\right) \frac{\bar{n}^{\mu }}{2}+\left( F\cdot \bar{n%
}\right) \frac{n^{\mu }}{2}+F_{\bot }^{\mu },
\end{equation}%
where $F_{\bot }$ denotes the components transverse to the lightlike
vectors : $\left( F_{\bot }\cdot n\right) =\left( F_{\bot }\cdot \bar{n}%
\right) =0$. Similarly, one can also write a decomposition 
\begin{equation}
F^{\mu }=\left( F\cdot \omega \right) \omega ^{\mu }+F_{\top }^{\mu },
\end{equation}%
where $F_{\top }$ denotes the component which is orthogonal to the velocity $%
\omega $: $\left( \omega \cdot F_{\top }\right) =0$.

The amplitudes defined in Eq.(\ref{Ame}) can be parametrised as%
\begin{equation}
\mathcal{A}_{\chi _{c0}\rightarrow V\gamma }=~\left( \epsilon _{V}^{\ast
}\cdot \epsilon _{\gamma }^{\ast }\right) ~M_{0}~A_{0V}^{\bot },
\label{def:A0}
\end{equation}%
\begin{equation}
\mathcal{A}_{\chi _{c1}\rightarrow V\gamma }=~i\varepsilon \lbrack \epsilon
_{\chi },\epsilon _{\gamma }^{\ast },p,q]~(\epsilon _{V}^{\ast }\cdot \omega
)\frac{m_{V}}{M_{1}^{2}}~A_{1V}^{\Vert }+i\varepsilon \lbrack \epsilon
_{V}^{\ast },\epsilon _{\gamma }^{\ast },p,q]~(\epsilon _{\chi }\cdot q)%
\frac{1}{M_{1}^{2}}~A_{1V}^{\bot },  \label{def:A1}
\end{equation}%
\begin{align}
\mathcal{A}_{\chi _{c2}\rightarrow V\gamma }& =\epsilon _{\chi }^{\alpha
\beta }q_{\alpha }\left( \epsilon _{\gamma }^{\ast }\right) _{\beta
}(\epsilon _{V}^{\ast }\cdot \omega )~\frac{m_{V}}{M_{2}}~A_{2V}^{\Vert
}+~\epsilon _{\chi }^{\alpha \beta }q_{\alpha }q_{\beta }\left( \epsilon
_{V}^{\ast }\cdot \epsilon _{\gamma }^{\ast }\right) \frac{1}{M_{2}}%
~A_{2V}^{\bot }  \notag \\
& +\epsilon _{\chi }^{\alpha \beta }\left\{ (\epsilon _{\gamma }^{\ast
})^{\alpha }(\epsilon _{V}^{\ast })^{\beta }+(\epsilon _{\gamma }^{\ast
})^{\alpha }q^{\beta }(\epsilon _{V}^{\ast }\cdot q)\frac{(pP)}{(pq)^{2}}%
-q_{\alpha }q_{\beta }\left( \epsilon _{V}^{\ast }\cdot \epsilon _{\gamma
}^{\ast }\right) \frac{1}{2E_{\gamma }^{2}}\right\} \frac{E_{\gamma }^{2}}{%
M_{2}}~T_{2V}^{\bot }.  \label{def:A2}
\end{align}%
where $\epsilon _{\chi }$ and $\epsilon _{V}^{\ast }$ denote polarisation
vectors (or tensors in the case of $\chi_{c2}$) of the initial and final mesons, respectively. Furthermore,  the amplitudes $A_{iV}^{\Vert }$
and $A_{iV}^{\bot }\ (T_{2V}^{\bot })$ correspond to the longitudinal and
transverse vector meson $V$, respectively.  In Eq.(\ref{def:A1}) the following  notation  has been used
\bea
i\varepsilon \lbrack \epsilon_{V}^{\ast },\epsilon _{\gamma }^{\ast },p,q]\equiv i\varepsilon_{\alpha_1\alpha_2\alpha_3\alpha_4}
(\epsilon_{V}^{\ast })^{\alpha_1}(\epsilon _{\gamma }^{\ast })^{\alpha_2}p^{\alpha_3} q^{\alpha_4}. 
\eea

The definitions of the amplitudes
in Eqs.(\ref{def:A0})-(\ref{def:A2}) are chosen in such a way that they are
dimensionless and can be associated with the corresponding helicity
amplitudes \ $\chi _{cJ}(\lambda _{\chi })\rightarrow V(\lambda _{V})\gamma
(\lambda _{\gamma })$. The simple analysis allows one to establish the
following  correspondence 
\begin{equation}
A_{0V}^{\bot }:\chi _{c0}\rightarrow V(\lambda _{V}=\pm 1)\gamma (\lambda
_{\gamma }=\pm 1),
\end{equation}%
\begin{eqnarray}
A_{1V}^{\bot } &:&\chi _{c1}(\lambda _{\chi }=0)\rightarrow V(\lambda
_{V}=\pm 1)\gamma (\lambda _{\gamma }=\pm 1), \\
~\ A_{1V}^{\Vert } &:&\chi _{c1}(\lambda _{\chi }=\pm 1)\rightarrow
V(\lambda _{V}=0)\gamma (\lambda _{\gamma }=\pm 1),
\end{eqnarray}%
\begin{eqnarray}
A_{2V}^{\bot } &:&\chi _{c2}(\lambda _{\chi }=0)\rightarrow V(\lambda
_{V}=\pm 1)\gamma (\lambda _{\gamma }=\pm 1), \\
~A_{2V}^{\Vert } &:&~\ \chi _{c2}(\lambda _{\chi }=\pm 1)\rightarrow
V(\lambda _{V}=0)\gamma (\lambda _{\gamma }=\pm 1), \\
T_{2V}^{\bot } &:&\ \chi _{c2}(\lambda _{\chi }=\pm 2)\rightarrow V(\lambda
_{V}=\mp 1)\gamma (\lambda _{\gamma }=\pm 1).
\end{eqnarray}

The expressions for the corresponding decay rates read
\begin{equation}
\Gamma \lbrack \chi _{c0}\rightarrow V\gamma ]=\frac{E_{\gamma }}{4\pi }%
\left\vert A_{0V}^{\bot }\right\vert ^{2}.
\end{equation}%
\begin{equation}
\Gamma \lbrack \chi _{c1}\rightarrow V\gamma ]=\frac{1}{12\pi }\frac{%
E_{\gamma }^{5}}{M_{1}^{4}}\left\{ |A_{1V}^{\Vert }|^{2}+|A_{1V}^{\bot
}|^{2}\right\} .
\end{equation}%
\begin{equation}
\Gamma \lbrack \chi _{c2}\rightarrow V\gamma ]=\frac{1}{40\pi }\frac{%
E_{\gamma }^{5}}{M_{2}^{4}}\left( \left\vert A_{2V}^{\Vert }\right\vert ^{2}+%
\frac{4}{3}\left\vert A_{2V}^{\bot }\right\vert ^{2}+2\left\vert
T_{2V}^{\bot }\right\vert ^{2}\right) .
\end{equation}

\subsection{The decay amplitudes with longitudinal light meson $\lambda _{V}=0$ }
 
 The QCD factorisation  framework implies that the decay amplitude can
be computed by expanding with respect to the following small parameters: $v$
the relative velocity of the heavy quarks and $\lambda \sim \sqrt{\Lambda /m}
$ where $\Lambda $ is the soft scale of order $\Lambda _{QCD}$. The
expansion with respect to $v$ can be carried out in the framework of NRQCD,
while the expansion with respect to $\lambda $ can be carried out within the
soft-collinear factorisation framework (SCET). The hard annihilation
mechanism is described by the appropriate partonic configuration in
accordance with the symmetries of QCD. The corresponding subprocess is
associated with the particles with large momenta $p_{i}^{2}\sim m^{2}\gg
\Lambda ^{2}$ and can be computed systematically within perturbative QCD.
The long distance contributions are associated with the matrix elements of
the operator defined in NRQCD and SCET. The relative order of  such a
configuration can be estimated  from the power counting defined within
these effective theories.

\begin{figure}[ptb]
\centering
\includegraphics[width=5.2478in]{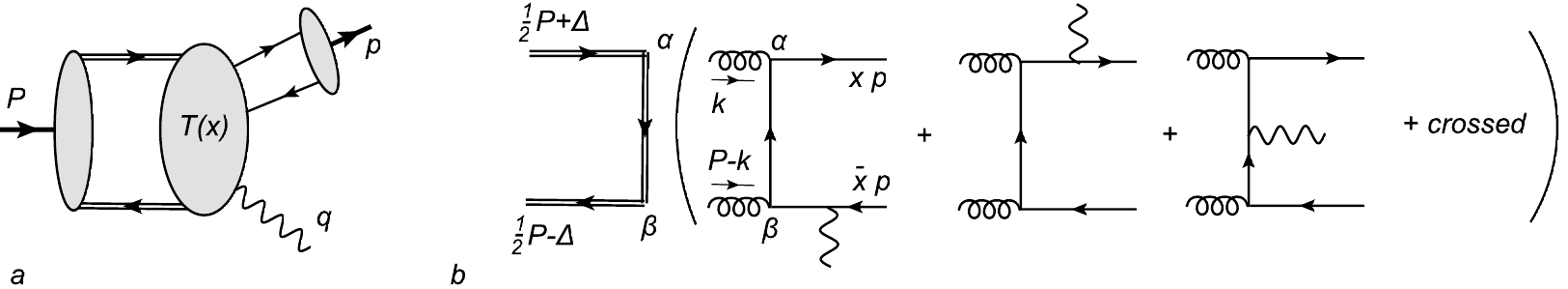}
\caption{The diagrams describing $(a)$ the hard
annihilation mechanism and  $(b)$ one-loop leading-order contribution
to the hard kernel $T(x)$. Here we assume that the gluon lines are attached
to the light quark lines to the vertices with the same indices 
$\alpha $ and $\beta $.}
\label{aiivdiagrams}
\end{figure}

In this case of the longitudinal  meson $V_{\Vert }$ the leading-order
contribution  is given by the diagrams in Fig.\ref{aiivdiagrams} . 
Two blobs in this figure correspond to the nonperturbative matrix
elements associated with the soft (NRQCD) and collinear (SCET) sectors 
of the effective theory.  The factorised analytical expressions for the
corresponding amplitudes can be presented in the following form%
\begin{equation}
~~i\varepsilon \lbrack \epsilon _{\chi },\epsilon _{\gamma }^{\ast
},p,q](\epsilon _{V}^{\ast }\cdot \omega )\frac{m_{V}}{M_{1}^{2}}%
A_{1V}^{\Vert }=i\left\langle \mathcal{O}(^{3}P_{1})\right\rangle 2(\epsilon
_{V}^{\ast }\cdot \omega )\sqrt{2}f_{V}~m_{V}Q_{V}\int_{0}^{1}dx~\phi
_{\Vert }^{V}(x)D_{1h}(x),  \label{A1IIV}
\end{equation}%
\begin{equation}
\epsilon _{\chi }^{\alpha \beta }q_{\alpha }\left( \epsilon _{\gamma }^{\ast
}\right) _{\beta }(\epsilon _{V}^{\ast }\cdot \omega )~A_{2V}^{\Vert
}=i\left\langle \mathcal{O}(^{3}P_{2})\right\rangle ~2(\epsilon _{V}^{\ast
}\cdot \omega )\sqrt{2}f_{V}~m_{V}Q_{V}\int_{0}^{1}dx~\phi _{\Vert
}^{V}(x)D_{2h}(x),  \label{A2IIV}
\end{equation}%
where $D_{ih}(x)$ denote an expression for the hard subdiagrams.  The
function  $\phi _{\Vert }^{V}(x)$ denotes the vector meson distribution
amplitude (DA) which is defined by the collinear matrix element,  
variable $x$ denotes the collinear fraction of the quark and  
 $f_{V}$ denotes the meson decay constant.  A more detailed
description of these matrix elements and the properties of DAs can be found
in  Appendix~A. The factor $Q_{V}$ in Eq.(\ref{A1IIV}) denotes an appropriate
combination of the quark charges 
\begin{equation}
Q_{\rho ^{0}}=\frac{1}{2}(e_{u}-e_{d})=\frac{1}{2},~Q_{\omega }=\frac{1}{2}%
(e_{u}+e_{d})=\frac{1}{6}~,~Q_{\phi }=e_s=-\frac{1}{3}.\ \ 
\end{equation}

The NRQCD soft matrix elements are defined as 
\begin{equation}
\left\langle 0\right\vert ~\mathcal{O}^{\beta }(^{3}P_{1})~\left\vert \chi
_{c1}(\omega )\right\rangle =\epsilon _{\chi }^{\beta }~i\left\langle 
\mathcal{O}(^{3}P_{1})\right\rangle ,  \label{def:O3P1}
\end{equation}%
\begin{equation}
\left\langle 0\right\vert ~\mathcal{O}^{\alpha \beta }(^{3}P_{2})~\left\vert
\chi _{c2}(\omega )\right\rangle =\epsilon _{\chi }^{\alpha \beta
}~i\left\langle \mathcal{O}(^{3}P_{2})\right\rangle ,  \label{def:O3P2}
\end{equation}%
where we imply the following operators 
\begin{equation}
\mathcal{O}^{\beta }(^{3}P_{1})=\frac{1}{2\sqrt{2}}~\chi _{\omega }^{\dag }%
\overleftrightarrow{D}_{\top }^{\alpha }\left( \frac{-i}{2}\right) \left[
\gamma _{\top }^{\alpha },\gamma _{\top }^{\beta }\right] \gamma _{5}\psi
_{\omega },
\end{equation}%
\begin{equation}
\mathcal{O}^{\alpha \beta }(^{3}P_{2})=\chi _{\omega }^{\dag }\left( -\frac{i%
}{2}\right) \overleftrightarrow{D}_{\top }^{(\alpha }\gamma _{\top }^{\beta
)}\psi _{\omega },
\end{equation}%
where $(\alpha ,\beta )$ denotes the symmetrical traceless tensor. These
operators are constructed from the quark $\psi _{\omega }$ and antiquark $%
\chi _{\omega }^{\dag }$ four-component spinor fields satisfying $\NEG%
{\omega}\psi _{\omega }=\psi _{\omega }$, $\NEG{\omega}\chi _{\omega }=-\chi
_{\omega }$. The constants on the \textit{rhs }of Eqs.(\ref{def:O3P1}) and (%
\ref{def:O3P2}) are related to the value of the charmonium wave functions at
the origin. To leading order in $v$ they read 
\begin{equation}
\left\langle \mathcal{O}(^{3}P_{J})\right\rangle \simeq \sqrt{2N_{c}}\sqrt{%
2M_{J}}\sqrt{\frac{3}{4\pi }}R_{21}^{\prime }(0),  \label{<O3PJ>}
\end{equation}%
where $R_{21}^{\prime }(0)$ is the derivative of the quarkonium radial wave
function.

With the given definitions of the nonperturbative matrix elements the
analytical expressions for the hard subdiagram $D_{ih}(x)$ in Eqs.(\ref{A1IIV}) and (\ref{A2IIV}) reads (Feynman gauge is implied) 
\begin{align}
D_{ih}(x)& =ie\alpha _{s}^{2}\frac{N_{c}^{2}-1}{4N_{c}^{2}}\int \frac{d^{4}k
}{(2\pi )^{4}}~\frac{1}{\left[ k^{2}\right] \left[ \left( k-P\right) ^{2}
\right] }  \notag \\
& ~\ \ \ \ \ \ \ \ \ \ \ \ \ \ \ \ \ \ \ \ \ \ \ \ \ \ \ \ \ \ \text{Tr}
\left[ \mathcal{P}_{iQ}^{\mu }\left( \frac{1}{2m}\left\{ \gamma _{\mu },{
D}_{Q}\right\} +{D}_{Q}^{\prime \mu }\right) \right] \text{Tr}[\mathcal{P
}_{V}~D_{q}].  \label{Dh:def}
\end{align}%
This expression is divided as  the product of  two gluon
propagators and two traces associated with heavy and light quark lines. The
notations $\mathcal{P}_{iQ}^{\mu }$ and $\mathcal{P}_{V}$ are used for
projectors on the heavy and light meson states 
\begin{equation}
\mathcal{P}_{1Q}^{\mu }=\sqrt{2}\epsilon _{\chi }^{\nu }\frac{1}{16}(1+\Dsl{%
\omega})\left[ \gamma _{\top }^{\mu },\gamma _{\top }^{\nu }\right] \gamma
_{5},~~\mathcal{P}_{2Q}^{\mu }=\frac{1}{8}\epsilon _{\chi }^{\mu \nu }(1+%
\Dsl{\omega})\gamma _{\top }^{\nu }~,~~\mathcal{P}_{V}=\frac{1}{8m}~\Dsl{p}.
\label{projectors}
\end{equation}%
The expressions for the \ functions of $\hat{D}_{Q}$ $\ $and $\hat{D}%
_{Q}^{\prime \mu }$ are generated by the heavy quark subdiagram, see Fig.\ref%
{aiivdiagrams} and read%
\begin{equation}
D_{Q}=\frac{\gamma ^{\beta }(m\Dsl{\omega} -\Dsl{k}+m)\gamma ^{\alpha }}{\left[
k^{2}-2m(k\omega )\right] },
\end{equation}%
\begin{equation}
D_{Q}^{\prime \mu }=\frac{1}{\left[ k^{2}-2m(k\omega )\right] }\left\{
\gamma ^{\beta }\gamma ^{\mu }\gamma ^{\alpha }+2k^{\mu }\frac{\gamma
^{\beta }(m\Dsl{\omega} -\Dsl{k}+m)\gamma ^{\alpha }}{\left[ k^{2}-2m(k\omega )\right] }%
\right\} .
\end{equation}%
The set of the light quark diagrams is given by the sum of the three subdiagrams
 as shown in Fig.\ref{aiivdiagrams} 
\begin{eqnarray}
D_{q} &=&\frac{\gamma ^{\alpha }(\ps_1+\ks)\gamma ^{\beta }(-\qs-\ps_2)
\epsilon_{\gamma }^{\ast }}{\left[ \left( k+p_1\right) ^{2}\right] \left[ (p_2+q)^{2}\right] }
+\frac{\gamma ^{\beta }(-\ks-\ps_2-\qs)\gamma ^{\alpha }(-\qs-\ps_2)\epsilon _{\gamma }^{\ast }}
{\left[ \left( k+p_2+q\right) ^{2}\right] \left[ (p_2+q)^{2}\right] }  \notag \\
&&+\frac{\gamma ^{\alpha }(\ps_1+\ks)\epsilon _{\gamma }^{\ast }(\ps_1+\qs+\ks)\gamma
^{\beta }}{\left[ \left( xp+k\right) ^{2}\right] \left[ (p_1+q+k)^{2}\right] }%
+\frac{\gamma ^{\beta }(-\ks-\ps_2-\qs)\epsilon _{\gamma }^{\ast }(-\ks-\ps_2)\gamma ^{\alpha }}{\left[ \left( k+p_2+q\right) ^{2}\right]
 \left[ (k+p_2)^{2}\right] }  \notag \\
&&+\frac{\epsilon _{\gamma }^{\ast }(\ps_1+\qs)\gamma ^{\alpha }(\ps_1+\qs+\ks)\gamma
^{\beta }}{\left[ (p_1+q)^{2}\right] \left[ (p_1+q+k)^{2}\right] }
+\frac{\epsilon _{\gamma }^{\ast }(\ps_1+\qs)\gamma ^{\beta }(-\ks-\ps_2)\gamma
^{\alpha }}{\left[ (p_1+q)^{2}\right] \left[ (k+p_2)^{2}\right] },\ 
\label{Dq:def}
\end{eqnarray}%
where we assume
\begin{equation}
p_{1}=x p,~p_{2}=\bar{x}p,
\end{equation}
and  $\bar{x}\equiv 1-x$. All propagators in the square brackets $%
[\ldots ]$ imply the standard Feynman prescription $+i\varepsilon $.

The sum of the one-loop diagrams must be IR finite because this is the leading-order contribution. Assuming that $k$ is hard $k\sim m$ one can easily see
that 
\begin{equation}
D_{h}(x)\sim \alpha _{s}^{2}(\mu _{h})\mathcal{O}(1),
\end{equation}%
where we assume the scaling behaviour with respect to small dimensionless
parameters $v$ and $\lambda $. Therefore the total scaling behaviour of such a
contribution is associated with the scaling of the operators in the
effective theory, 
\begin{equation}
\mathcal{O}^{\beta }(^{3}P_{1})\sim v^{4},~\ \bar{q}_c \nbs q_c \sim
\lambda ^{4}, 
\label{scaling}
\end{equation}
where $q_c$ denotes the collinear quark field. 
Hence the leading-order contribution to the amplitudes $A_{iV}^{\Vert }$ can
be estimated as 
\begin{equation}
A_{iV}^{\Vert }\sim \alpha _{s}^{2}(\mu _{h})v^{4}\lambda ^{2},
\label{AIIV-singlet}
\end{equation}%
where we take into account that the external collinear hadronic state gives
a factor $\lambda ^{-2}$ due to the normalisation.

The hard factorisation can be violated if there is an overlap with the ultrasoft 
region when  one of the gluons has ultrasoft momentum of order $mv^{2}$. 
Such a contribution can be associated with the colour-octet mechanism.
Performing the expansion of the expression for $D_{q}$ with respect to small $k\sim mv^2$  one finds that
 potentially dangerous terms cancel:
\bea
\text{Tr}[\mathcal{P}_{V}D_{q}]_{us}&\simeq& \frac{1}{8m}\frac{2p^{\alpha }%
\text{Tr}\left[ \ps\gamma ^{\beta }(-\qs)\epsilon _{\gamma }^{\ast }\right] }{%
\left[ 2\left( kp\right) \right] \left[ 2\bar{x}(pq)\right] }+\frac{1}{8m}%
\frac{2p^{\alpha }\text{Tr}\left[ \ps\epsilon _{\gamma }^{\ast }\qs\gamma
^{\beta }\right] }{\left[ 2\left( kp\right) \right] \left[ 2x(pq)\right] }
\nonumber \\ & &
+\frac{1}{8m}\frac{-2p^{\alpha }\text{Tr}\left[ \ps\gamma ^{\beta
}(-\qs)\epsilon _{\gamma }^{\ast }\right] }{\left[ 2\bar{x}(pq)\right] \left[
2(kp)\right] }+\frac{1}{8m}\frac{-2p^{\alpha }\text{Tr}\left[ \ps\epsilon
_{\gamma }^{\ast }\qs\gamma ^{\beta }\right] }{\left[ 2\left( pk\right) \right]
\left[ 2x(pq)\right] }=0.\ 
\eea
This cancellation is a consequence of the colour neutrality of the outgoing
quark-antiquark pair. Similarly one can see that the contribution from the region
where $P-k\sim mv^{2}$ also vanishes.  This allows us to conclude that the
colour-octet mechanism is suppressed by a power of $v^{2}$ comparing to the
contribution of the hard region. Therefore the loop integrals in Eqs.(\ref%
{Dh:def}) can only have IR divergencies in the individual diagrams and these
singularities must cancel in the sum of all diagrams.

Performing the necessary calculations we obtain
\begin{equation}
A_{1V}^{\Vert }=-i\left\langle \mathcal{O}(^{3}P_{1})\right\rangle ~\frac{%
~f_{V}M_{1}^{2}}{m^{6}}Q_{V}\sqrt{4\pi \alpha }\alpha _{s}^{2}(\mu _{h})~%
\frac{N_{c}^{2}-1}{2N_{c}^{2}}\int_{0}^{1}dx~\phi _{V}^{\Vert }(x)T_{1}(x),
\label{A1IIV-res}
\end{equation}%
\begin{equation}
A_{2V}^{\Vert }=-i\left\langle \mathcal{O}(^{3}P_{2})\right\rangle ~\frac{%
f_{V}M_{2}}{m^{5}}Q_{V}\sqrt{4\pi \alpha }\alpha _{s}^{2}(\mu _{h})~\frac{%
N_{c}^{2}-1}{2N_{c}}\frac{1}{2\sqrt{2}}\int_{0}^{1}dx~\phi _{V}^{\Vert
}(x)T_{2}(x),  \label{A2IIV-res}
\end{equation}%
with the following hard kernels%
\begin{equation}
T_{i}(x)=\func{Re}T_{i}\left( x\right) +i\func{Im}T_{i}(x),
\end{equation}%
\begin{eqnarray*}
\func{Re}T_{1}\left( x\right) &=&-\frac{\pi ^{2}}{12}\frac{1}{\bar{x}^{3}}-%
\frac{x}{4\bar{x}^{3}}\ln ^{2}2+\left( -\frac{1}{\bar{x}^{2}}-\frac{1}{4\bar{%
x}}-\frac{3}{4x}\right) \ln 2+\left( -\frac{3}{4\bar{x}}+\frac{1}{4x}-\frac{1%
}{2x-1}\right) \ln \bar{x} \\
&&+\frac{x}{\bar{x}^{3}}\ln x\ln \bar{x}+\left( -\frac{1}{2\bar{x}^{2}}+%
\frac{3}{4\bar{x}}-\frac{1}{4x}+\frac{1}{2x-1}\right) \ln x-\frac{3x}{4\bar{x%
}^{3}}\ln ^{2}x-\frac{x}{2\bar{x}^{3}}\ln x\ln 2
\end{eqnarray*}%
\begin{equation}
-\frac{1}{2}\frac{x}{\bar{x}^{3}}\left( \text{Li}\left[ 1-\frac{1}{2x}\right]
+\text{Li}\left[ 1-2x\right] +\text{Li}\left[ 1-x\right] +\text{Li}\left[ -%
\bar{x}/x\right] -\text{Li}\left[ 2x-1\right] \right) +(x\rightarrow \bar{x}%
),
\end{equation}%
\begin{equation}
\func{Im}T_{1}\left( x\right) =\frac{\pi }{4x\bar{x}^{3}}~\left( \bar{x}%
(1+x(2x-1))+2x^{2}\ln [x]\right) +(x\rightarrow \bar{x}),
\end{equation}%
\begin{equation*}
\func{Re}T_{2}(x)=-\left( \frac{\pi ^{2}}{3}+\ln ^{2}2\right) \frac{(1+\bar{x%
})(1+x)}{2\bar{x}^{4}}+\ln 2\left\{ \frac{8}{\bar{x}^{3}}-\frac{10}{\bar{x}%
^{2}}-\frac{4}{\bar{x}x}+\frac{8\bar{x}}{(2x-1)^{2}}\right\}
\end{equation*}%
\begin{eqnarray*}
&&+\ln 2\ln x\frac{(1+x)(2+3x)}{\bar{x}^{4}}+\ln x\left\{ \frac{4}{\bar{x}%
^{3}}-\frac{3}{\bar{x}^{2}}+\frac{3}{2\bar{x}}-\frac{1}{2x}-\frac{2}{%
(2x-1)^{2}}\right\} +\ln ^{2}x\frac{(1+x)(5x-2)}{2\bar{x}^{4}} \\
&&+2\ln x\ln \bar{x}\frac{(1+\bar{x})(1+x)}{\bar{x}^{4}}-2\ln \bar{x}\left\{ 
\frac{1}{\bar{x}^{2}}+\frac{3}{\bar{x}}+\frac{1}{x}+\frac{1}{(2x-1)^{2}}+%
\frac{1}{2x-1}\right\}
\end{eqnarray*}%
\begin{equation}
-\frac{(1+\bar{x})(1+x)}{\bar{x}^{4}}\left\{ \text{Li}\left[ 1-\frac{1}{2x}%
\right] +\text{Li}\left[ 1-2x\right] +\text{Li}\left[ 1-\frac{1}{x}\right] +%
\text{Li}\left[ \bar{x}\right] +\text{Li}\left[ 2x-1\right] \right\}
+(x\rightarrow \bar{x}),
\end{equation}%
\begin{equation}
\func{Im}T_{2}=\frac{2\pi }{x\bar{x}^{4}}\left(
1-5x+7x^{2}-5x^{3}+2x^{4}-x^{2}(1+x)\ln x~\right) +(x\rightarrow \bar{x}),
\end{equation}%
where 
\begin{equation}
\text{Li}\left[ x\right] \equiv \text{Li}_{2}\left[ x\right] =-\int_{0}^{x}dt%
\frac{\ln \bar{t}}{t},
\end{equation}%
is the Spence function. These hard kernels have singular endpoint behaviour
\begin{equation}
\func{Re}T_{i}\left( x\right) \overset{x\rightarrow 1}{\sim }\frac{\ln \bar{x}}{\bar{x}},\ \func{Im}T_{i}\left( x\right) \overset{x\rightarrow 1}{\sim }\frac{1}{\bar{x}}
\end{equation}%
but these singularities are compensated by the endpoint suppression  of the DA  $\phi _{\Vert }^{V}(x\rightarrow 1)\sim \bar{x}$ therefore the convolution
integrals in Eqs.(\ref{A1IIV-res}) and (\ref{A2IIV-res}) are well defined.
We also assume the hard scale $\mu _{h}\sim m$ in the argument of the QCD
running coupling.

Above we obtained the well-defined formula for the longitudinal
amplitudes. However these contributions at leading order are already
suppressed by a small factor $\alpha _{s}^{2}(\mu _{h})$. 

 This could  reduce the value of  colour-singlet contribution in comparison with the colour-octet one which potentially can  be of order $\alpha_s(\mu_h)$.  
 In the realistic case  when  $mv^2\sim \Lambda$  the  colour-octet  matrix element is nonperturbative;  therefore its  computation is difficult and can 
 be done only within a model-dependent framework.  The other possibility is to perform the analysis of the leading  colour-octet correction in the 
 Coulomb limit when $mv^{2}\gg $\ $\Lambda $. In  this case the scales $v$ and $\lambda $ are well separated  $v\gg  \lambda$, and the 
 charmonium state can be considered as a perturbative  Coulomb state.  Such a situation allows one to establish  the  well-defined scaling behaviour 
 with respect to the small parameters $v$ and $\lambda$.   The details of our  analysis  can be found in Appendix~B.   We obtain the scaling behaviour 
\begin{equation}
\left[ A_{1V}^{\Vert }\right] _{oct}\sim \alpha _{s}(\mu _{h})\alpha
_{s}(\mu _{us})~v^{6}\lambda ^{2}, \label{AIIoct-1-sectionII}
\end{equation}%
where we introduced the ultrasoft scale $\mu_{us}\sim mv^2$.
Therefore  the ratio of octet to singlet amplitudes in the Coulomb limit behaves as   $\left[
A_{1V}^{\Vert }\right] _{oct}/\left[ A_{1V}^{\Vert }\right] _{sing}\sim
\alpha _{s}(\mu _{us})v^{2}/\alpha _{s}(\mu _{h})$. 

\begin{figure}[h]
\centering
\includegraphics[width=5.0 in]
{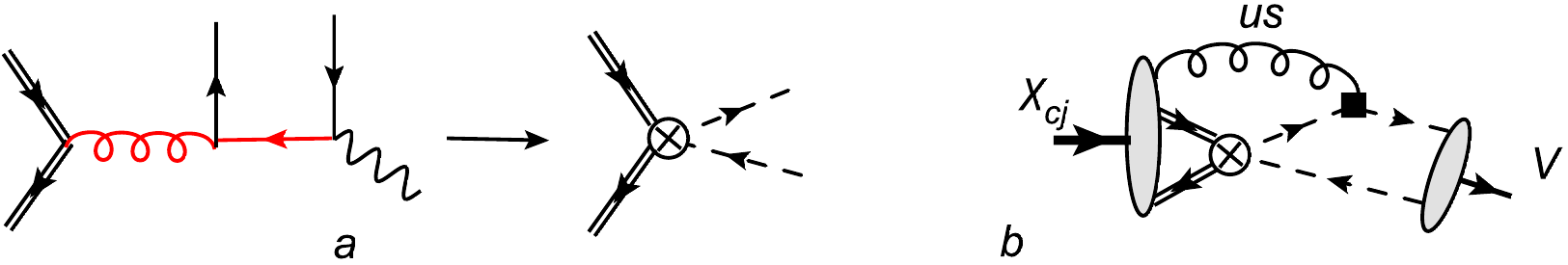}
\caption{  $(a)$ The illustration of the hard factorisation which gives  the colour-octet operator. The red lines denote the hard propagators. The resulting operator has colour structure $\chi^\dagger_\omega T^a \psi_\omega \, \bar q_{hc} T^a q_{hc}$.  The dashed lines denote the hard-collinear fields $q_{hc}$.  $(b)$  The diagram which illustrates the formal factorisation of the  hard-collinear modes.  The black square denotes the subleading SCET vertex.   The corresponding  ultrasoft  gluon is absorbed into the NRQCD  matrix element with a colour-octet operator. }
\label{hard-fact}
\end{figure}

For the real charmonium   $v\sim \lambda$ and  one can perform only the hard factorisation which  gives the power $\alpha_s(\mu_h)$ 
and a four-quark  operator constructed from the heavy quark-antiquark fields (colour-octet operator $\mathcal{O}^8({}^3S_1)$  in NRQCD,  see Eq.(\ref{O8-3S1}) ) 
 and hard-collinear fields, see  Fig.\ref{hard-fact}$(a)$. 
 One can assume  that the corresponding matrix  element describes the soft overlap of the  heavy and light mesons wave functions. 
 Because $v\sim \lambda$  we assume that  ultrasoft fields in NRQCD and nonperturbative soft fields in SCET  with $k_s\sim \Lambda$ 
 coincide. In order to estimate the power behaviour of the colour-octet matrix element  we integrate over  the hard-collinear  modes in SCET, see Fig.\ref{hard-fact}$(b)$.   
  The interactions of the soft and hard-collinear fields in this case  remain the same as in the Coulomb limit  and can be described
   by the same subleading interactions in SCET suppressed by a small scale $\lambda$ (not $v$ as in the Coulomb limit). 
   Therefore instead of  powers of velocity $v$ one obtains the same powers of $\lambda$, with the difference that  we now assume $v\sim \lambda$.  
The colour-octet operator  $\mathcal{O}^8({}^3S_1)$ overlaps with the $\chi_{cJ}$  states at order $\mathcal{O}(v)$ and this is also the same as in the Coulomb limit. 
   This allows us to conclude that the  scaling  behaviour  of the  colour-octet matrix element can be 
  obtained from  Eq.(\ref{AIIoct-1-sectionII})  assuming that   $\alpha_s(\mu_{us})\sim 1$. i.e. 
\begin{equation}
 { \left[  A_{1V}^{\Vert } \right] _{oct} }\, /\,{ \left[ A_{1V}^{\Vert }\right] _{sing} }
 \sim{v^{2}}/{\alpha _{s}(\mu _{h})} \sim {\lambda^{2}}/{\alpha _{s}(\mu _{h})},
 \label{oct2sing}
\end{equation}
For the charm quark the numerical values $v^{2}\simeq 0.3$ and $\alpha _{s}(2m_{c}^{2})=0.29$ and therefore numerically,  the
relative size of these contributions is of order one. Hence  we can conclude that in this case  the colour-octet contribution
may potentially  be  important.   If this is true then we must also observe this from the numerical estimates 
when we take into account only the colour-singlet  matrix elements.  We  will perform such a numerical study  in Sec.~\ref{phen}.  

\subsection{Decay amplitudes with transverse light meson.}

\begin{figure}[ptb]
\centering
\includegraphics[width=4.8767in]{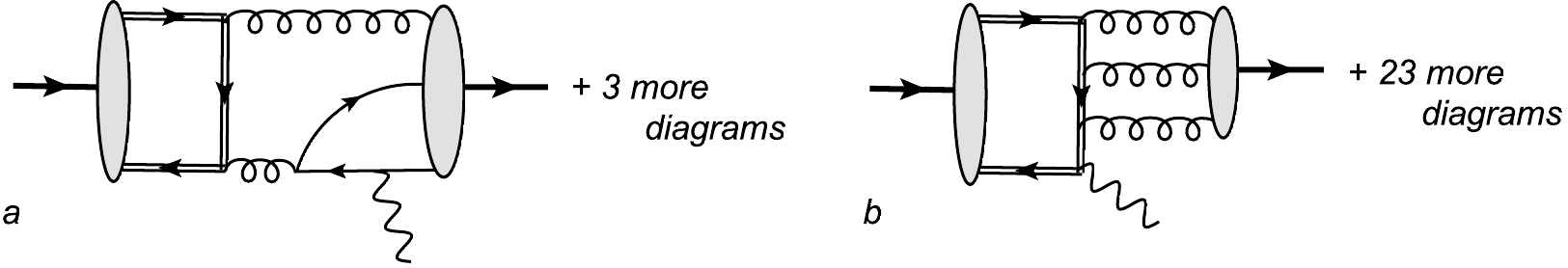}
\caption{The leading order diagrams which
provide contribution to the amplitude $A_{V}^{\bot }$. }
\label{avperpdiagrams}
\end{figure}

For the
final state with transverse meson one has to consider  the twist-3
distribution amplitudes. There are two different possibilities: the photon
is emitted from the light quark or from the heavy quark lines.  Examples
of the corresponding diagrams are shown in Fig.\ref{avperpdiagrams}. 
The contributions with the
three gluon DAs are possible only for the isosinglet mesons $\omega $ and $%
\phi $.  All these contributions are of order $\alpha _{s}$ and the
corresponding hard kernels are described by tree diagrams. Their 
 analytical expressions are quite lengthy and we will not write
them explicitly. The corresponding calculations are similar to the one described
above and yield  the following expressions
\begin{eqnarray}
A_{0V}^{\bot } &=&i\left\langle \mathcal{O}(^{3}P_{0})\right\rangle \frac{%
f_{V}m_{V}}{M_{0}m^{4}}\sqrt{4\pi \alpha }\frac{\pi \alpha _{s}}{N_{c}}\int
D\alpha _{i}\left\{ \frac{Q_{V}}{\sqrt{6}}\frac{\left( 2+\alpha _{3}\right) 
}{\alpha _{1}\alpha _{2}\alpha _{3}^{2}}\left( \frac{\alpha _{1}-\alpha _{2}%
}{1-\alpha _{3}}V(\alpha _{i})+A(\alpha _{i})\right) \right.  \notag \\
&&~\ \ \ \ \ \ \ \ \ \ \ \ \ \ \ \ \ \ \ \ \ \ \ \ \ \ \ \ \ \ \ \ \ \ \ \ \
\ \ \ \ \ \ \ \ \ \left. +~\delta _{I0}~e_{Q}2\sqrt{6}~\frac{G(\alpha _{i})}{%
\alpha _{1}\alpha _{2}\alpha _{3}}\frac{2+\alpha _{1}}{\alpha _{1}(1-\alpha
_{1})}\right\} ,  \label{A0perpV}
\end{eqnarray}%
\begin{eqnarray}
A_{1V}^{\bot } &=&i\left\langle \mathcal{O}(^{3}P_{1})\right\rangle ~\frac{%
~f_{V}m_{V}}{m^{5}}\frac{M_{1}^{2}}{m^{2}}\sqrt{4\pi \alpha }\frac{\pi
\alpha _{s}}{N_{c}}\int D\alpha _{i}\left\{ \delta _{I0}~9e_{Q}~\frac{%
G(\alpha _{i})}{\alpha _{1}\alpha _{2}\alpha _{3}^{2}}\right.  \notag \\
&&~\ \ \ \ \ \ \ \ \left. -\frac{Q_{V}}{4}\left( \frac{\alpha _{1}-\alpha
_{2}}{\alpha _{1}\alpha _{2}\alpha _{3}^{2}}V(\alpha _{i})+\frac{1-\alpha
_{3}}{\alpha _{1}\alpha _{2}\alpha _{3}^{2}}A(\alpha _{i})\right) \right\} ,
\label{A1perpV}
\end{eqnarray}%
\begin{eqnarray}
~A_{2V}^{\bot } &=&i\left\langle \mathcal{O}(^{3}P_{2})\right\rangle \frac{%
f_{V}m_{V}}{m^{5}}\frac{M_{2}}{m}\sqrt{4\pi \alpha }\frac{\pi \alpha _{s}}{%
N_{c}}\int D\alpha _{i}\left\{ -\delta _{I0}3\sqrt{2}e_{Q}~\frac{G(\alpha
_{i})}{\alpha _{1}\alpha _{2}\alpha _{3}^{2}}\right.  \notag \\
&&~\ \ \ \ \ \left. -\frac{Q_{V}}{4\sqrt{2}}\left( \frac{\alpha _{1}-\alpha
_{2}}{\alpha _{1}\alpha _{2}\alpha _{3}^{2}}V(\alpha _{i})+\frac{1-\alpha
_{3}}{\alpha _{1}\alpha _{2}\alpha _{3}^{2}}A(\alpha _{i})\right) \right\} ,
\label{A2perpV}
\end{eqnarray}%
\begin{eqnarray}
~T_{2V}^{\bot } &=&i\left\langle \mathcal{O}(^{3}P_{2})\right\rangle \frac{%
f_{V}m_{V}}{E_{\gamma }m^{4}}\frac{M_{2}}{E_{\gamma }}\sqrt{4\pi \alpha }%
\frac{\pi \alpha _{s}}{N_{c}}\int D\alpha _{i}\left\{ -\delta _{I0}3\sqrt{2}%
e_{Q}\frac{G(\alpha _{i})}{\alpha _{1}\alpha _{2}\alpha _{3}^{2}}\frac{1}{%
1-\alpha _{3}}\right.  \notag \\
&&~\ \ \ \ \ \ \ \ \ \ \ \ \ \ \ \ \ \ \ \left. +\frac{Q_{V}}{2\sqrt{2}}%
\frac{1}{\alpha _{1}\alpha _{2}\alpha _{3}^{2}}\left( \frac{\alpha
_{2}-\alpha _{1}}{1-\alpha _{3}}V(\alpha _{i})+A(\alpha _{i})\right)
\right\} .  \label{T2perpV}
\end{eqnarray}
In these formulas it is implied   that  $e_{Q}$ is the electric charge of the heavy quark, the symbol $\delta _{I0}$
specifies the contribution which exists only for the meson states with 
isospin $I=0$. We also used the shorthand notation for the collinear integrals 
\begin{equation}
\int D\alpha _{i}~f(\alpha _{i})=\int_{0}^{1}d\alpha _{1}\int_{0}^{1}d\alpha
_{2}\int_{0}^{1}d\alpha _{3}~\delta (1-\alpha _{1}-\alpha _{2}-\alpha
_{3})~f(\alpha _{1},\alpha _{2},\alpha _{3}),  \label{Da}
\end{equation}%

The matrix element for the scalar charmonia reads 
\begin{equation}
\left\langle 0\right\vert \frac{-1}{\sqrt{3}}~\chi _{\omega }^{\dag }\left( 
\frac{-i}{2}\right) \overleftrightarrow{D}_{\top }^{\alpha }\gamma _{\top
}^{\alpha }\psi _{\omega }~\left\vert \chi _{c0}\right\rangle =i\left\langle 
\mathcal{O}(^{3}P_{0})\right\rangle .
\end{equation}
where the constant $\left\langle \mathcal{O}(^{3}P_{0})\right\rangle $ is
given by expression in Eq.(\ref{<O3PJ>}).

In order to describe the overlap with the transverse light mesons we need three-particle twist-3
DAs  $V(\alpha _{i})$, $A(\alpha _{i})$  and $G(\alpha _{i})$.  
A  detailed description of these nonperturbative functions is given in  
Appendix~A.  The properties of  these functions allow one to conclude that the convolution integrals in
Eqs.(\ref{A0perpV})-(\ref{T2perpV}) are well defined.

The NRQCD matrix element in  Eqs. (\ref{A0perpV})-(\ref{T2perpV})  is of order $v^4$,  the twist-3 operators in the  collinear sector are of order $\lambda ^{6}$. 
Therefore one finds 
\begin{equation}
A_{iV}^{\bot }\sim T_{iV}^{\bot }\sim \alpha _{s}(\mu _{h})v^{4}\lambda ^{4}.
\label{TiVsingle}
\end{equation}%
Hence these amplitudes are suppressed by a factor $\lambda ^{2}$ and enhanced by 
$\alpha _{s}$ compared to the longitudinal ones.

An appropriate   contribution  from the  colour-octet  operator  is considered   in Appendix~B.  We  obtain
that in the Coulomb limit the contribution from the colour-octet matrix element behaves as 
\bea
\left[ A_{1V}^{\bot }\right]_{oct}\sim \alpha _{s}(\mu _{h})\alpha_{s}(\mu _{us})\lambda^4 v^4.
\label{Atw3-oct-clmb}
\eea
Hence following  the same arguments as in the previous subsection and extrapolating this result to the real charmonium with  $v\sim \lambda $ we expect that
this contribution is of the same order as the singlet one
 \bea
\left[ A_{1V}^{\bot }\right]_{oct}/\left[ A_{1V}^{\bot }\right]_{sing}\sim 1.
\label{Atw3-oct}
\eea
 Therefore an estimate made with the help of the singlet contribution potentially may have a large uncertainty due to the colour-octet contribution.

\section{Phenomenology}
\label{phen}

In this section we study  numerical estimates using only  contributions of the colour-singlet operators. Our aim is to
understand how well one can  describe the charmonium  decays  in this case  using reliable  estimates for various hadronic parameters. 

In our numerical  calculations  we are using  the following  nonperturbative input.
For the $c$-quark mass we take the value $m_{c}=1.5\ $GeV, for the charmonium
states $M_{0}=3.42\, $GeV, $M_{1}=3.51\, $GeV and $M_{2}=3.56\, $GeV and for the
light mesons $m_{\rho }=775\, $MeV, $m_{\omega }=783\, $MeV, $m_{\phi }=1019\, $MeV.
For the derivative of the radial wave function we use the value from
Ref.\cite{Eichten:1995ch}  computed for the Buchm\"uller-Tye potential%
\begin{equation}
|R_{21}^{\prime }(0)|^{2}=0.75\, \text{GeV}^{5}.
\end{equation}

For the description of the  light meson matrix elements we use the
following values of the decay constants%
\begin{equation}
f_{\rho }=221\, \text{MeV},~\ \ f_{\omega }=198\, \text{MeV},~\ \ f_{\phi }=161\, \text{MeV},
\label{fVMeV}
\end{equation}%
which are defined according to Eq.(\ref{DA:def}). 

The estimates for the  vector meson DAs  have been 
studied in many works, see e.g. Refs.\cite{Ball:1996tb, Ball:1998sk, Ball:1998ff, Ball:2007zt}. 
The leading-order DA is
described by the model with one parameter 
\begin{equation}
\phi _{V}(x,\mu )=6x\bar{x}\left\{ 1+a_{2}^{V}(\mu
)C_{2}^{3/2}(2x-1)\right\} ,
\end{equation}%
For the coefficients $a_{2}^{V}$ we use the values  from  Ref. \cite{Ball:2007zt} 
\begin{equation}
a_{2}^{\rho }=a_{2}^{\omega }=0.15\pm 0.07,~a_{2}^{\phi }=0.18\pm 0.08,
\label{a2inSR}
\end{equation}
where all parameters are given at the scale $\mu =1\, $GeV. 

Using the explicit expressions for the coefficient functions $T_{i}(x)$ one can easily 
obtain the values for the convolution integrals 
\begin{equation}
\int_{0}^{1}dx~\phi _{V}(x)T_{i}(x)=A_{i}+a_{2}^{V}(\mu _{h})~B_{i},
\end{equation}%
with 
\begin{equation}
A_{i}=\int_{0}^{1}dx~6x\bar{x}~T_{i}(x),~\ B_{i}=\int_{0}^{1}dx~6x\bar{x}%
C_{2}^{3/2}(2x-1)~T_{i}(x).
\end{equation}%
Their numerical values read 
\begin{eqnarray}
A_{1} &=&1.32+5.46i,~~\ A_{2}=-12.71+6.01i~,~ \\
\ B_{1} &=&7.00+4.79i,~\ B_{2}=9.61+6.12i.
\end{eqnarray}

The transverse amplitudes depend on the twist-3 DAs defined in 
Eqs.(\ref{def:Atw3})-(\ref{Gtld}) in Appendix~A.  For the quark-gluon DAs  we  use the   models given in Eq.(\ref{tw3-AV-mod})
which has been  suggested in Refs.\cite{Ball:1996tb, Ball:1998sk, Ball:1998ff} .
The nonperturbative parameters which enter in these formulas have been  taken from Ref. \cite{Ball:2007zt}
at the scale $\mu =1\, $GeV: 
\begin{eqnarray}
\rho ~\text{and }\omega \text{-mesons} &\text{:~\ }&\zeta _{3}=0.030\pm
0.010,~\ \omega _{3}^{A}=-3.0\pm 1.4,~\ \omega _{3}^{V}=5.0\pm 2.4,
\label{tw3in} \\
\phi \text{-meson} &\text{:~\ }&\zeta _{3}=0.024\pm 0.008,~\ \omega
_{3}^{A}=-2.6\pm 1.3,~\ \omega _{3}^{V}=5.3\pm 3.0~.
\end{eqnarray}%
In the following estimates we neglect  the small difference for the $\phi $-meson and consider as a first guess  that all 
parameters are constrained  only by the values in  Eq.(\ref{tw3in}).

For the isosinglet mesons $\omega $ and $\phi $ we  have an additional contribution
from the three gluon DAs. Taking into account the conformal expansion and mixing with the quark operators
(see details in the Appendix) we use for them the  models given in Eq.(\ref{tw3-G-mod}).
 We assume that the value  of the corresponding local  matrix element is small because we expect a very
small pure gluon component of the meson wave function at low scale $\mu =1$GeV.
 Hence the  constant  $\omega _{3}^{G}$  must be 
 much smaller then the corresponding constants  of the quark gluon operators in Eq.(\ref{tw3norm}) or  
\begin{equation}
|\omega _{3}^{G}(\mu =1\text{GeV})|\ll 1.
\end{equation}
We consider $\omega _{3}^{G}$ as a free parameter and try to estimate its value from the
comparison with the data.  The QCD evolution of  all  the twist-3 parameters is described 
in  Appendix~A. 

The numerical results also depend on the choice of the hard scale $\mu
_{h}$. In the following calculations  it is fixed  to be  $\mu
_{h}^2=2m_{c}^{2}$ if it is not  written otherwise.  For the total decay
rates we used the data from Ref.\cite{Olive:2016xmw}: $\Gamma _{tot}[\chi
_{cJ}]=\{10.5,0.84,1.93\}$MeV for $J=0,1,2$, respectively.   Finally, in the following calculations we use the NLO QCD
 coupling which has the value $\alpha _{s}(2m_{c}^{2})=0.290$ 
  %(or $\alpha _{s}(m_{\tau }^{2})=0.317$).

We start our discussion from the description of the branching rations $Br\left[
\chi _{cJ}\rightarrow V_{\Vert }\gamma \right] $ because  these observables are largest  for  $\chi _{c1}$ decays. 
The obtained results are given in Table \ref{BrXc1-VII}.
\begin{table}    
\centering 
\begin{tabular}{|c|c|c|c|}\hline
& $\gamma \rho $ & $\gamma \omega $ & $\gamma \phi $ \\ \hline 
$\chi _{c1}\rightarrow V_{\Vert }\gamma $ & 
$\begin{array}{cc}
{} \\[-2mm]
153.1_{-16.7-70.5}^{+18.2+103.7}
\end{array}
$ & $13.6_{-1.5-6.3}^{+1.6+9.2}$ & $31.3_{-3.8-14.5}^{+4.2+21.4}$ \\[2mm] 
\hline
$ \chi _{c2}\rightarrow V_{\Vert }\gamma$ & 
$\begin{array}{cc}
{} \\[-2mm]
2.11_{-0.08-0.9}^{+0.09+1.3}
\end{array}
$ & $0.19_{-0.007-0.08}^{+0.008+0.12}$ & $0.41_{-0.02-0.18}^{+0.02+0.26}$ \\[2mm] \hline
\end{tabular}%
\caption{The obtained values for the $Br\left[ \chi _{c1,2}\rightarrow V_{\Vert }\gamma \right]$  in units of $10^{-6}$. }
\label{BrXc1-VII} 
\end{table}
The first  error  shows the sensitivity to the value of the parameter $a_{2}^{V}$ within the intervals given in Eq.(\ref{a2inSR}). 
 The second error  shows the dependence on $\mu _{h}$ within  the interval $m_{c}<\mu _{h}<2m_{c}.$  
 The corresponding errors are large  because the decay rates  are proportional to the fourth  power of the QCD coupling 
  $Br\left[ \chi _{c1}\rightarrow V_{\Vert }\gamma  \right]  \sim \alpha _{s}^{4}(\mu _{h})$. 
 With  the given estimate for $R'_{21}(0)$ the obtained  values  for $Br\left[ \chi
_{c1}\rightarrow \rho _{\Vert }\gamma \right] $ and $Br\left[ \chi
_{c1}\rightarrow \phi _{\Vert }\gamma \right] $ are quite reliable although the obtained numbers lie somewhat below/above  the experimental results 
given in Table~\ref{dataXc1}.  However, the
estimate for $Br\left[ \chi _{c1}\rightarrow \omega _{\Vert }\gamma \right] $ is about a factor of four smaller than the experimental value. One can also consider, for instance, 
the ratio $Br\left[ \chi _{c1}\rightarrow \omega _{\Vert }\gamma \right] /Br%
\left[ \chi _{c1}\rightarrow \rho _{\Vert }\gamma \right] $ in which  the
normalisation ambiguities cancel.  If one assumes that the dominant
contribution to the amplitudes arises from the terms associated with $u$ and 
$d$-quark components of the electromagnetic current in Eq.(\ref{Ame}) then 
using  $SU(2)$ symmetry one finds that this ratio must be  
\begin{equation}
\frac{Br\left[ \chi _{c1}\rightarrow \omega _{\Vert }\gamma \right] }
{Br\left[ \chi _{c1}\rightarrow \rho _{\Vert }\gamma \right] }\simeq \frac{1}{9},
\label{ratio}
\end{equation}%
However  the experimental value for this ratio  is $0.28$ which is about a factor of three larger. 
The large difference   indicates that the assumption about the small
contribution of the  $c$-quark components in Eq.(\ref{Jem}) is not valid.  
However the  longitudinal amplitude in which the photon is emitted from a heavy quark  $A^\Vert_{Q}$ can only be  generated from  a colour-octet operator.      
 The basic idea is that one gluon in the diagram in Fig.\ref{2g-hard}$(c)$ is ultrasoft  while the other two are hard-collinear. These gluons create a light quark-antiquark pair in the octet state which after interaction   with the ultasoft  gluon  becomes colorless, see also Fig.~\ref{col-oct-d2}.  In Appendix~B  we consider such colour-octet  matrix element in the Coulomb limit. 
We obtain that such contribution behaves as 
\begin{equation}
A_{Q}^{\Vert }\sim \alpha _{s}(\mu _{h})\alpha _{s}(\mu _{us})~\lambda ^{2}v^{7}.
\end{equation}%
where we used the fact that $\alpha _{s}(\mu\sim mv)\sim v$,   provides one more  power of $v$. 
 Following  the same arguments as in Sec.\ref{hard} we expect that  for  real  charmonium with $v\sim \lambda$  this amplitude can be estimated as 
\begin{equation}
A_{Q}^{\Vert }\sim \alpha _{s}(\mu _{h})\lambda ^{2}v^{7}.
\end{equation}%
Therefore  this amplitude is only suppressed by a factor $v$ compared to the  colour-octet  amplitude $[A^\Vert_{1V}]_{oct}$ in Eq.(\ref{oct2sing}). 
We assume that  the interference of this amplitude  with the larger amplitudes  $A^\Vert_{1V}$ and  $[A_V^{\Vert}]_{oct}$  can be responsible for the large  ratio in Eq.(\ref{ratio}).
Perhaps, these contributions strongly enhance the decay rate  of the $\omega \gamma$ channel.  One reason could be  that the used value of
$R'_{21}(0)$ is somewhat large and therefore this numerically  enhances  the singlet contribution.  In case the realistic value  $R'_{21}(0)$  is smaller, then the relative effect of the colour-octet contribution in the $\rho\gamma$ channel must  be larger  in order to  describe the data.  In any case  it seems that the  contribution  of the colour-octet amplitude $A_{Q}^{\Vert }$ must play an important role in the correct description of  the $\omega _{\Vert }\gamma $ decay mode.

  The computed  branching ratios for the state $\chi _{c2}$ are substantially smaller and they easily satisfy  the experimental
constraints. When comparing our results to Ref.\cite{Gao:2006bc}, we note that our  estimates for $\chi _{c1}$  shown in Table \ref{BrXc1-VII}  are  about an order of magnitude larger than theirs,  while the results for the $\chi _{c2}$  decays are in good agreement. 

We next consider the  decay rates with transversely polarised mesons in the final
state.  The numerical evaluation with 
arbitrary twist-3 DA parameters (at scale $\mu =1$GeV) gives
\begin{equation}
\Gamma _{\rho }^{\bot }=222.4~\zeta _{3}^{2}(-9.82+4.78\omega
_{3}^{A}+3.31\omega _{3}^{V})^{2},  \label{Grho-theor}
\end{equation}%
\begin{equation}
\Gamma _{\omega }^{\bot }=181.2~\zeta _{3}^{2}(-3.3+1.4~\omega
_{3}^{A}+8.3~\omega _{3}^{V}+735.6~\omega _{3}^{G})^{2},  \label{GTw-theor}
\end{equation}%
\begin{equation}
\Gamma _{\phi }^{\bot }=168.7~\zeta _{3}^{2}(6.5-3.3~\omega
_{3}^{A}+2.9~\omega _{3}^{V}+733.6~\omega _{3}^{G})^{2},  \label{GTphi-theor}
\end{equation}%
 One can  observe that the coefficients in front of the gluon parameter $\omega _{3}^{G}$
are always quite large, hence  even the relatively small values of $\omega
_{3}^{G}$ can produce a significant numerical impact. Formally, the large coefficient in Eqs.(\ref{GTw-theor}) and (\ref{GTphi-theor}) arises due to  the  large normalization 
 of the twist-3 gluon DA  in Eq.(\ref{tw3-G-mod}).  

\begin{figure}[ptb]
\centering
\includegraphics[width=2.5313 in]
{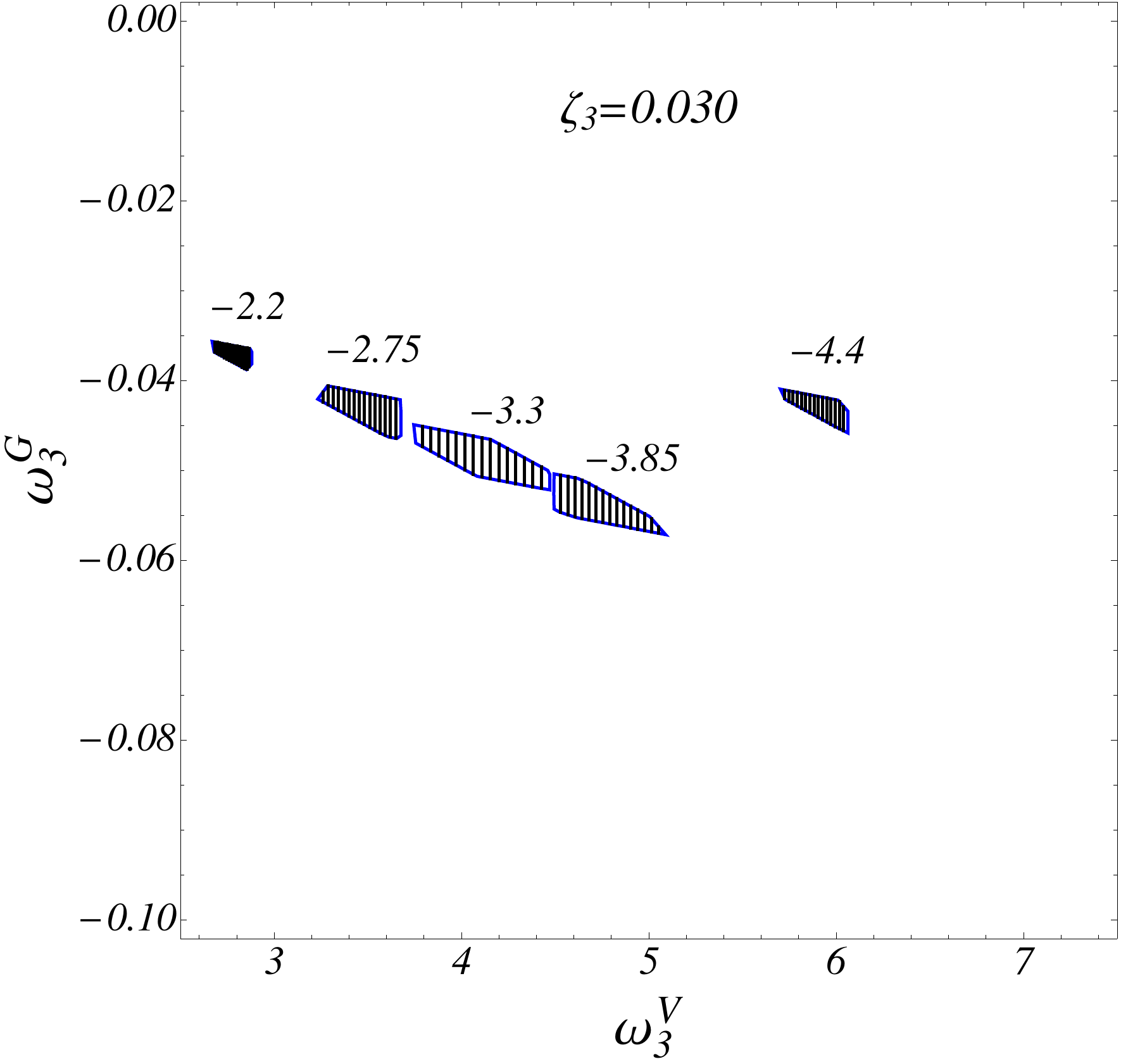}
\ \ \
\includegraphics[width=2.5313 in]
{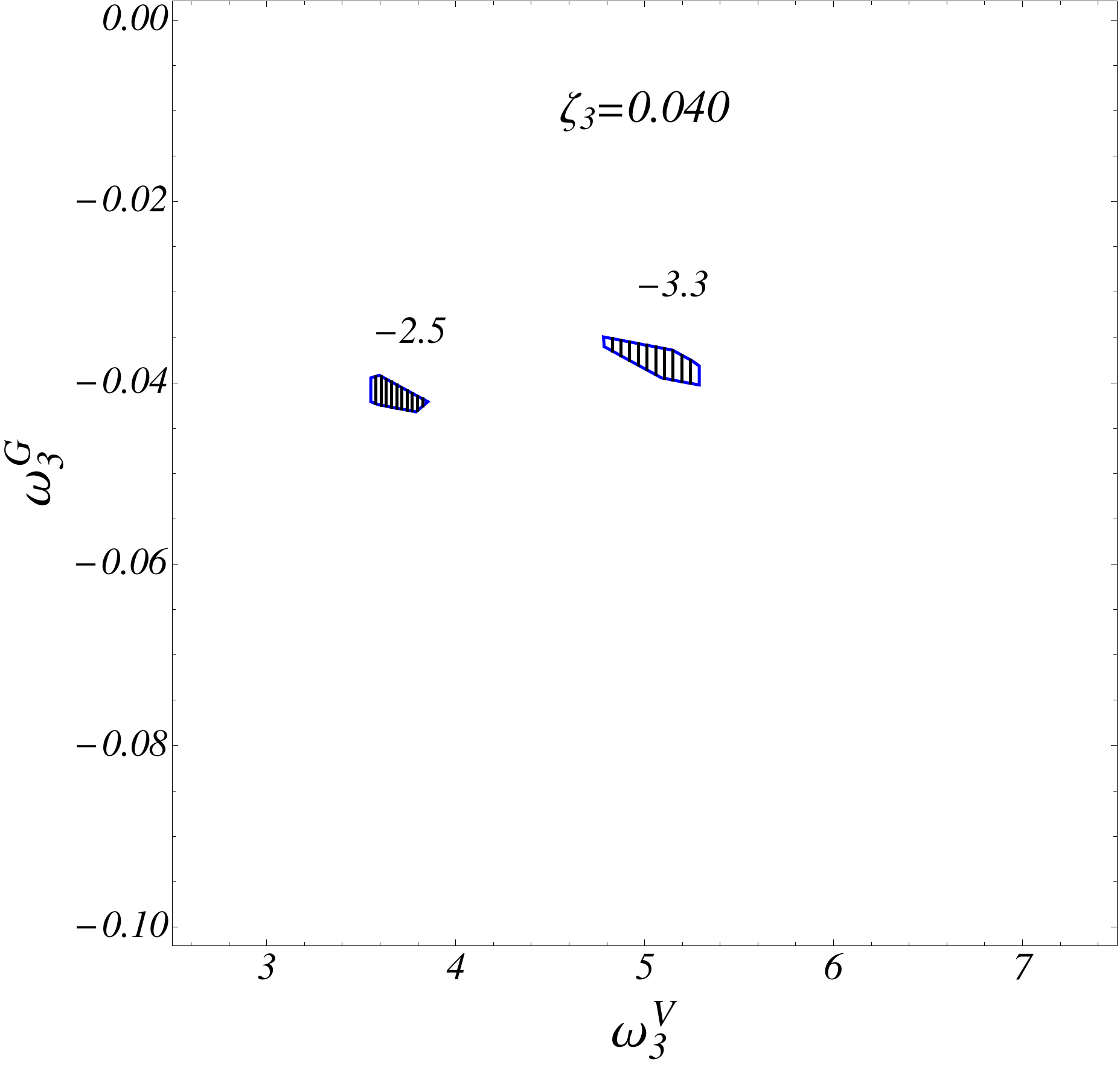}
\caption{The numerical restrictions for the DA
parameters $\omega _{3}^{V}$ and $\protect\omega _{3}^{G}$ at fixed $\zeta _{3}$ and $\omega _{3}^{A}$ values (indicated on the allowed regions in  both figures). }
\label{plot-regions-xi}
\end{figure}

The existing data do not allow us to fix the  DA
parameters in Eqs.(\ref{Grho-theor})-(\ref{GTphi-theor}) in order to
unambiguously  predict  the  decay widths for $\chi _{c0,2}$.  From the
numerical estimates we observe that there are different solutions which
describe the data for $\chi _{c1}\rightarrow V_\perp\gamma$ decays within the experimental error bars
and at the same time provide very different  estimates for the $\chi _{c0,2}$
decays.  In order to illustrate this let us consider a few examples. It is
convenient to fix the parameters $\zeta _{3}$ and $\omega _{3}^{A}$ in
accordance with the estimates in Eq.(\ref{tw3in})  and to study the possible
restrictions on the  two remaining constants $\omega _{3}^{V}=5.0\pm 2.4$ and $
\left\vert \omega _{3}^{G}\right\vert <0.1$. The latter inequality  must be
considered as an assumption.  For simplicity, for all final mesons $V$ we imply  the  value $a_{2}=0.22$ in the
longitudinal amplitudes of $\chi _{c2}$ decays. In the following we
consider  two fixed values: $\zeta _{3}=0.03$ and $\zeta _{3}=0.04$.  The
different possible solutions for these values of $\zeta _{3}$ for a few
different values of $\omega _{3}^{A}$ are shown in Fig.\ref{plot-regions-xi}.
In the given
regions we can describe the data for $\chi _{c1}$  satisfying  the
experimental constraints for the $\chi _{c0,2}$ decays.  We observe
that the strongest restrictions on the DA parameters are provided by the
data for $\chi _{c1}$ decay. The restrictions on $\chi _{c2}$ decay rates are
relatively   weak  but  also allow us to get some constraints.  The computed values
for the decay widths of $\chi _{c0}$ are always so  small that they do not contradict  the
experimental restrictions.  

 In Table~\ref{XcJperp}  we show  examples of the different  values of parameters and  corresponding  results for  the branching fractions.
\begin{table}[th]
\centering
\begin{tabular}{|c|c|c|c||c|c|c||c|c|c||c|c|c|}
\hline
\multicolumn{4}{|c||}{} &\multicolumn{3}{|c||} {$\chi_{c1}\rightarrow V_\perp \gamma$ } & 
\multicolumn{3}{|c||} {$\chi_{c2}\rightarrow V \gamma$ }   
& \multicolumn{3}{|c|} {$\chi_{c0}\rightarrow V \gamma$ }  \\ \hline
 $\zeta_3$& $\omega_3^{A}$ & $\omega_3^{V} $&$ \omega_3^{G} $  & $\rho$ & $\omega$ & $\phi$ & $\rho$ & $\omega$ & $\phi $ & $\rho$ & $\omega$ & $\phi$ \\ \hline
 0.03 & -2.2 & 2.8 & -0.037 & 29.6 & 20.8 & 4.8 & 3.4 & 0.18 & 2.6 & 2.0 & 0.17 & 0.66 \\ \hline
0.03& -4.4 & 5.9 & -0.043 & 30.0 & 13.9 & 8.5 & 17.2 & 3.7 & 6.5 & 0.40 & 0.05 & 0.15 \\ \hline
0.04 & -2.5 & 3.7 &-0.041 & 39.2 & 13.5 & 6.5 & 7.1 & 0.40 &5.6 & 1.9 & 0.16 & 0.54 \\ \hline
0.04 &-3.4 & 5.1 &-0.038 & 33.2 & 14.2 & 6.2 & 16.3 & 3.6 & 6.50 & 0.62 & 0.09 & 0.23 \\ \hline
\end{tabular}
\caption{The examples of the DA parameters and the corresponding values of  branching  fractions in units of $10^{-6}$.}
\label{XcJperp}
\end{table}
Note that we found the  values of $\omega^{G}_3$ to be always negative,  within  this approach. 
We also observe  that the twist-3 gluon  contribution is critically important for a description of  decays  with   $\omega$ and $\phi$  mesons.  
From the numerical results we find that 
\begin{equation}
Br[\chi_{c1}\rightarrow\gamma\rho]>Br[\chi_{c2}\rightarrow\gamma\rho
]>Br[\chi_{c0}\rightarrow\gamma\rho],
\end{equation}
and%
\begin{equation}
Br[\chi_{c2}\rightarrow\gamma\rho]>Br[\chi_{c2}\rightarrow\gamma\phi
]>Br[\chi_{c2}\rightarrow\gamma\omega].
\end{equation}
One can also see that there are  solutions for which 
\begin{equation}
Br[\chi_{c2}\rightarrow\gamma\phi]\geq Br[\chi_{c1}\rightarrow\gamma\phi].
\end{equation}

The obtained results allow  us to conclude that a further progress in our understanding of the
radiative decays would come  from a measurement of  the branching for the $Br[\chi_{c2}\rightarrow \gamma \rho ]$. It will also be interesting to carry out
the corresponding helicity analysis. The  estimates  in Table~\ref{XcJperp} show that for the 
$\chi _{c2}$ decays the transverse branching  fraction can be much larger the the longitudinal one; see Eq.(\ref{BrXc1-VII}). The dominant
contribution is provided by the  amplitude $T_{2\rho }^{\bot }$ which
describes the decay of the charmonium state with helicity $\lambda =\pm 2$. Hence a prediction for the $Br[\chi_{c2}\rightarrow \gamma \rho ]$ obtained  only from  the 
$\rho_L\gamma$ contribution can strongly underestimate the realistic  value of the branching.  
 Notice that for $\chi _{c1}$ decay the situation is opposite: the dominant amplitude  is $A_{1\rho }^{\Vert }$ describing the decay of $\chi _{c1}(\lambda =\pm 1)$.

 We also see that the used non-perturbative input and DAs models potentially  allows one to describe the existing data for the transverse decays without  the colour-octet contributions. 
If such a scenario is realised  then  the radiative decays  can be used for a study of the twist-3  wave functions of the vector mesons.  The future data on these decays will be helpful
in order  to reduce the ambiguities  in the presented  description and would allow us to further constrain    the role of the various contributions.

\section{Conclusions}
\label{conc}

In the present paper we studied the radiative decays of the $P$-wave charmonia $\chi_{cJ}\rightarrow \gamma V$ with $V=\rho,\omega,\phi$.    
The QCD factorisation has been used in a systematic way: NRQCD  in the heavy meson sector  and a collinear expansion  in order to describe the overlap with  light  mesons in the final state. The  colour-singlet  contributions to all  helicity amplitudes  have been computed using the light-cone distribution amplitudes of twist-2 and twist-3. All obtained expressions are  well defined at least in the leading-order approximation.   The  colour-octet contributions  have  also been studied  in the Coulomb limit  in order to obtain their scaling behaviour.  

In order to understand the relevance of the  different  contributions we performed numerical estimates of the branching fractions using   the colour-singlet approximations for the decay amplitudes.   
 The various  models for the  vector meson distribution amplitudes which are available  in the literature   have been used 
 in our  numerical calculations.

  Our results indicate that  for the $\chi_{c1}\rightarrow \gamma V_\Vert $ decay the  colour-singlet contributions alone  do not allow to provide a  reliable description of { \it all} branching 
  fractions.  In particular, the branching fraction for  $\gamma \omega$ final state is about a factor four  smaller than the observed value.   We expect that  the colour-octet matrix elements  play   an important role in this case.  Qualitatively this conclusion also agrees with  the observation that the ratio of the octet to singlet amplitudes   behaves as $[A^\Vert_{1V}]_{octet}/[A^\Vert_{1V}]_{singlet}\sim  v^2/\alpha_s(\mu\sim m_c)$  which is not so small  for charmonia.  In addition, the   $\omega\gamma$ decay amplitude might  be enhanced by  the specific gluonic colour-octet operator  which can be responsible for  the large ratio of the branching fractions for $\omega$ and $\rho$ mesons.  We obtain that the corresponding  amplitude  is only suppressed  by a factor of $v$ in comparison with the amplitude   $[A^\Vert_{1V}]_{octet}$.   
  
 For  decays with a transverse meson in the final state  the  singlet and octet  operators contribute to the same order.  Using only the singlet contributions  we have shown that one can obtain a good description of the data  but  one needs to include the contributions with the  three-gluon distribution amplitude which is quite sizeable in size.  This observation may indicate a small color-octet contribution which would allow one to better  constrain  the  twist-3 DAs of the vector mesons.  We expect  that  future more precise measurements of the decay rates  $\chi_{c2}\rightarrow V \gamma$   will help to reduce the theoretical uncertainties and further clarify the  role of the various  contributions in these reactions.

% \section*{Aknowlegements}

\begin{appendix}
\numberwithin{equation}{section}
\section{Light-cone distribution amplitudes of the vector mesons}

Here we briefly describe the properties of various  distribution amplitudes  which have been used in our calculations. For simplicity we use here the 
light-cone gauge 
\bea
\bar{n}\cdot A(z)=0.
\eea
The  required two-particle light-cone matrix elements read
\bea
\left\langle V(p)\right\vert \bar{q} (\lambda _{1}\bar{n})\gamma_\mu q (\lambda _{2}\bar{n})\left\vert 0\right\rangle 
=(\epsilon_{V}^{\ast }\cdot \bar{n} )\frac{1}{2}n^\mu~\sqrt{2}f_{V}m_{V}\int_{0}^{1}dx~\phi_V^{\Vert}(x)e^{i\lambda _{1}x\left( p\bar{n}\right) +i\lambda _{2}\bar{x}\left( p\bar{n}\right) }
\nonumber \\ +
 (\epsilon_{V}^{\ast })_{\mu_\perp}  \sqrt{2} f_{V}m_{V} \int_{0}^{1}dx~g_V^v(x)e^{i\lambda _{1}x\left( p\bar{n}\right) +i\lambda _{2}\bar{x}\left( p\bar{n}\right) }
\label{DA:def}
\\
\left\langle V(p)\right\vert \bar{q} (\lambda _{1}\bar{n})\gamma_\mu \gamma_5 q (\lambda _{2}\bar{n})\left\vert 0\right\rangle 
=\frac12 i\varepsilon[\mu,\epsilon^*_V,p,\bar{n}]\frac12(\lambda_1-\lambda_2)  \sqrt{2} f_{V}m_{V}\int_{0}^{1}dx~g_V^a(x)e^{i\lambda _{1}x\left( p\bar{n}\right) +i\lambda _{2}\bar{x}\left( p\bar{n}\right) },
\eea
where we neglected the contributions of twist-4.
For  all operators in this paper we assume an appropriate flavour structure  
\bea
\left\langle \rho^0\right\vert(\bar{u}  u -\bar{d}d)\left\vert 0\right\rangle, \ \left\langle \omega\right\vert (\bar{u}  u+\bar{d}d)\left\vert 0\right\rangle, \
\left\langle \phi\right\vert \bar{s}  s\left\vert 0\right\rangle.
\eea

The required three-particle twist-3 matrix elements are given by  (recall that  $p_{-}\equiv (p\bar n) $)
\begin{equation}
\left\langle V(p)\right\vert \bar{q}(\lambda _{1}\bar{n})g\tilde{G
}_{\bar{n}\nu }(\lambda _{3}\bar{n})\nbs \gamma _{5}q(\lambda
_{2}\bar{n})\left\vert 0\right\rangle =-\sqrt{2}f_{V}m_{V}\left( \epsilon
_{V}^{\ast }\right) _{\nu }p_{-}^{2}~\mathcal{A}(\lambda _{i}p_{-}),
\label{def:Atw3}
\end{equation}%
\begin{equation}
\left\langle V(p)\right\vert \bar{q}(\lambda _{1}\bar{n})gG_{\bar{%
n}\nu }(\lambda _{2}\bar{n})\nbs q(\lambda _{3}\bar{n}%
)\left\vert 0\right\rangle =i\sqrt{2}f_{V}m_{V}\left( \epsilon _{V}^{\ast
}\right) _{\nu }p_{-}^{2}~\mathcal{V}(\lambda _{i}p_{-}),
\label{def:Vtw3}
\end{equation}%
\begin{equation}
\left\langle V(p)\right\vert ~gd^{abc}G_{\bar{n}\xi }^{a}(\lambda _{1}\bar{n}%
)G_{\bar{n} }^{b\xi}(\lambda _{2}\bar{n})G_{\bar{n}\nu }^{c}(\lambda _{3}%
\bar{n})\left\vert 0\right\rangle =i\sqrt{2}f_{V}m_{V}(\epsilon ^{\ast})_{\nu } p_{-}^{3}~\mathcal{G}(\lambda _{i}),  \label{G:def}
\end{equation}%
\begin{equation}
\left\langle V(p)\right\vert gd^{abc}G_{\bar{n}\xi }^{a}(\lambda _{1}\bar{n})%
\tilde{G}_{\bar{n} }^{b\xi}(\lambda _{2}\bar{n})\tilde{G}_{\bar{n}\nu
}^{c}(\lambda _{3}\bar{n})\left\vert 0\right\rangle =-i\sqrt{2}%
f_{V}m_{V}(\epsilon ^{\ast})_{\nu } p_{-}^{3}~\mathcal{\tilde{G}}(\lambda
_{i}),  \label{Gtld}
\end{equation}%
where  $d^{abc}$ is the fully symmetrical structure constant of the $SU(3)$ colour group and 
\begin{equation}
G_{\bar{n}\xi }^{a}(z)=\bar{n}^{\alpha }G_{\alpha \xi }^{a}(z),~\tilde{G}_{%
\bar{n}\xi }^{a}=\frac{1}{2}\varepsilon _{\alpha \xi \mu \nu }\bar{n}%
^{\alpha }G^{\mu \nu }(z).
\label{Gnbar}
\end{equation}%
The vector decay couplings $f_V$ can be obtained from the leptonic decays. Their explicit values are given in Eq.(\ref{fVMeV}).

All functions on the $rhs$ of Eqs. (\ref{def:Atw3})-(\ref{Gtld}) can be presented  as 
\begin{equation}
\mathcal{A(V,\mathcal{G, \tilde{G} } )}(\lambda _{i}p_{-})=\int D\alpha _{i}~A(V,G, \tilde{G})(\alpha
_{i})e^{i\left( p\bar{n}\right) (\lambda _{1}\alpha _{1}+\lambda _{2}\alpha
_{2}+\lambda _{3}\alpha _{3})},
\end{equation}%
with the integration measure $D\alpha _{i}$  defined in Eq.(\ref{Da}).

The properties of the DAs, except the pure gluonic ones,  have been  discussed in Refs.\cite{Ball:1996tb, Ball:1998sk, Ball:1998ff, Ball:2007zt}.
The  two-particle  functions are symmetrical with respect to exchange $x\leftrightarrow \bar x$ and normalised as 
\bea
\phi_V^{\Vert}(1-x)=\phi_V^{\Vert}(x), \ g^{v,a}_V(1-x)=g^{v,a}_V(x),  
\\  
    \int_0^1dx\ \phi_V^{\Vert}(x)= \int_0^1dx g^{v,a}_V(x)=1.
\eea
The  QCD equation of motions and operator identities allow one to express the twist-3 DAs $g^{v,a}_V$ in terms of  $\phi_V^{\Vert}$ and twist-3  functions $V$ and $A$, see e.g. 
\cite{Ball:1998sk}. Neglecting the three-particle contributions with $V$ and $A$ one has
\bea
g_{V}^v(x)=\frac{1}{2}\int_{0}^{x}dv\frac{\phi_V^\Vert(v)}{\bar{v}}+\frac{1}{2}\int%
_{x}^{1}dv\frac{\phi_V^\Vert(v)}{v},
\label{gv}
\\
g^{a}_V(x)=2\bar{x}\int_{0}^{x}dv\frac{\phi_V^\Vert(v)}{\bar{v}}+2x\int_{x}^{1}%
dv\frac{\phi_V^\Vert(v)}{v}.
\label{ga}
\eea
These expressions  are often referred to as the Wandzura-Wilczeck  approximation. 

The models for the  functions  $V$, $A$ have been  constructed using the conformal expansion, see details in Ref. \cite{Ball:1998sk}.   
Corresponding  functions   include the  contributions from local operators  with conformal spin  $j=7/2$ and $j=9/2$ and read
\begin{equation}
A(\alpha _{i})=360\zeta _{3}\alpha _{1}\alpha _{2}\alpha _{3}^{2}\left(
1+\omega _{3}^{A}\frac{1}{2}\left( 7\alpha _{3}-3\right) \right) ,\quad
V(\alpha _{i})=540\zeta _{3}\omega _{3}^{V}\alpha _{1}\alpha _{2}\alpha
_{3}^{2}(\alpha _{2}-\alpha _{1}),
\label{tw3-AV-mod}
\end{equation}%
The quark-gluon operators in our case  mix with the pure gluonic ones. Moreover, at the leading-logarithmic approximation such mixing is possible only for the operators with equal conformal spin.  Hence constructing the models for the  gluon  DAs  $G$ and $\tilde G$ we consider  only the contributions with the same conformal spin as  for the quark-gluon DAs.  From the definitions of the operators in Eqs.(\ref{G:def}) and (\ref{Gtld})  one can see that
\bea
G(\alpha_2, \alpha_1,\alpha_3)=G(\alpha_1, \alpha_2,\alpha_3), \ \   \tilde G(\alpha_2, \alpha_1,\alpha_3)=-\tilde G(\alpha_1, \alpha_2,\alpha_3).
\eea
Using this information  and following  the  arguments as in Ref.\cite{Ball:1998sk}  one obtains that the conformal expansion of the function $G$ and  $\tilde{G}$ is starting from the conformal spin $j=9/2$ and  $j=11/2$, respectively.  However as soon as contributions with the spin $j\geq11/2$ in $V$ and $A$ have been neglected  we assume that  $\tilde{G}(\beta _{i})$  can also be neglected.  Therefore
\begin{equation}
G(\alpha _{i})=5040~\zeta _{3}\omega _{3}^{G}\alpha _{1}^{2}\alpha
_{2}^{2}\alpha _{3}^{2},\quad  \tilde{G}(\alpha _{i})\simeq 0.  
\label{tw3-G-mod}
\end{equation}
The twist-3 DAs have the following normalization
\begin{equation}
\int D\alpha _{i}A(\alpha _{i})=\zeta _{3},~~\int D\alpha _{i}(\alpha
_{2}-\alpha _{1})V(\alpha _{i})=\zeta _{3}\omega _{3}^{V}\ ,~~\int D\alpha
_{i}G(\alpha _{i})=\zeta _{3}\omega _{3}^{G}.
\label{tw3norm}
\end{equation}%.
In general,  the parameters $\zeta_3$ and $\omega_3^{A,V,G}$  are different for the $\rho,\ \omega$ and $\phi$  mesons. 
For the sake of simplicity we neglect   this difference at low scale $\mu_0=1$GeV in our numerical calculations and assume
\bea
\left.\zeta_3(1\text{GeV})\right|_\rho=\left.\zeta_3(1\text{GeV})\right|_\omega=\left.\zeta_3(1\text{GeV})\right|_\phi,
\\
\left.\omega_3^{A,V}(1\text{GeV})\right|_\rho=\left.\omega_3^{A,V}(1\text{GeV})\right|_\omega=\left.\omega_3^{A,V}(1\text{GeV})\right|_\phi,
\\
\left.\omega_3^{G}(1\text{GeV})\right|_\omega=\left.\omega_3^{G}(1\text{GeV})\right|_\phi .
\eea
Therefore  we do not write an additional index associated with the mesonic state assuming that this information is  clear from the context.
\footnote{ Let us also mention  that the indices $V$ and $A$  for  $\omega_3^{A,V}$ are related with the Dirac structure of the quark-gluon operator.}

 All  distribution amplitude parameters  have to be evolved from the  low scale $\mu_0=1$GeV to some hard scale $\mu\sim m$.  The evolution of the twist-2  coupling  $a_2^V$
 is simple and well known, see e.g. Ref.\cite{Ball:1996tb}.   The evolution of the twist-3 DAs is more complicated and we take a closer look at it. 
 The evolution of the coupling   $\zeta_3$ (the lowest conformal spin $j=7/2$) is also simple and for all mesons one has, see e.g.  \cite{Ball:1998ff}
 \bea
 \zeta_3(\mu)=\left(  \frac{\alpha_{s}(\mu)}{\alpha_{s}(\mu_{0})}\right)  ^{{\gamma}_{00}/b} \zeta_3(\mu_0),
 \eea
 with the anomalous dimension 
 \begin{equation}
\gamma_{00}=-\frac{1}{3}C_{F}+3C_{A},
\end{equation}
and $b=\frac{11}{3}N_{c}-\frac{2}{3}n_{f}$. 
As the parameters  $\omega_3^{V,A}$   are mixing ($j=9/2$), we introduce 
\bea
\boldsymbol{\omega}_{3}(\mu)=
\left(
\begin{array}
[c]{c}%
\omega_{3}^{A} (\mu)\\
\omega_{3}^{V} (\mu)
\end{array}  \right).
\eea
For the $\rho$ meson one then has 
\bea
\rho\text{-meson}\quad
\boldsymbol{\omega}_3(\mu)
=\left(  \frac{\alpha_{s}(\mu)}{\alpha_{s}%
(\mu_{0})}\right)  ^{\hat{\gamma}_{(8)}/b}
\boldsymbol{\omega}_3 (\mu_0),
\label{Lg8rho}
\eea 
with  the $2\times 2$ anomalous dimension matrix  $\hat{\gamma}_{(8)}$ which will be given below. 

 The evolution of the corresponding parameters for the $\omega$ and $\phi$ 
is  more involved  because of mixing  with the gluon coupling $\omega_3^{G}$ and due to  the flavour  mixing. 
Performing the evolution of the   twist-3 parameters $\zeta _{3}$,  $\omega_3^{V,A,G}$  we assume that at  low scale $\mu=1$GeV,  $\omega \sim (u\bar u+d\bar d)$ and
 $\phi\sim s\bar s$ and therefore the matrix elements with  $s$- and $u(d)$-quarks vanish, respectively.  Our results for the evolved couplings read 
\bea
\omega\text{-meson }\quad \boldsymbol{\omega}_3(\mu)=\frac{1}{\sqrt{3}}\boldsymbol{\omega}_3^{(8)}(\mu)+\sqrt{\frac{2}{3}}\boldsymbol{\omega}_3^{(1)}(\mu),
\\
\phi\text{-meson }\quad \boldsymbol{\omega}_3(\mu)=-\sqrt{\frac{2}{3}}\boldsymbol{\omega}_3^{(8)}(\mu)+\frac{1}{\sqrt{3}}\boldsymbol{\omega}_3^{(1)}(\mu),
\eea
where
\bea
\boldsymbol{\omega}_3^{(8)}(\mu)=\left(  \frac{\alpha_{s}(\mu)}{\alpha_{s}%
(\mu_{0})}\right)  ^{\hat{\gamma}_{(8)}/b}
\boldsymbol{\omega}_3^{(8)} (\mu_0),\quad   
\left(
\begin{array}
[c]{c}%
\boldsymbol{\omega}_3^{(1)}(\mu)
\label{Lg8w38}\\
\omega_{3}^{G}(\mu)
\end{array}  \right) 
=\left(  \frac{\alpha_{s}(\mu)}{\alpha_{s}%
(\mu_{0})}\right)  ^{\hat{\gamma}_{(1)}/b}
\left(
\begin{array}
[c]{c}%
\boldsymbol{\omega}_3^{(1)}(\mu_0)\\
\omega_{3}^{G}(\mu_0)
\end{array}  \right) ,
\label{Lg1w1wG}
\eea
with the appropriate initial conditions
\bea
\omega\text{-meson }\quad   \boldsymbol{\omega}_3^{(8)} (\mu_0)=\frac{1}{\sqrt{3}} \boldsymbol{\omega}_3(\mu_0),\quad  \boldsymbol{\omega}_3^{(1)} (\mu_0)=\sqrt{\frac{2}{3}}\boldsymbol{\omega}_3(\mu_0),
\\ 
\phi\text{-meson }\quad   \boldsymbol{\omega}_3^{(8)} (\mu_0)=-\sqrt{\frac{2}{3}} \boldsymbol{\omega}_3(\mu_0),\quad  \boldsymbol{\omega}_3^{(1)} (\mu_0)=\frac{1}{\sqrt{3}}\boldsymbol{\omega}_3(\mu_0).
\eea
The anomalous dimension matrices in Eqs.(\ref{Lg8rho})  and (\ref{Lg1w1wG}) read
\begin{equation}
\hat{\gamma}_{(8)}=\left(
\begin{array}
[c]{cc}%
\gamma_{AA}-\gamma_{00} & \gamma_{AV}\\
\gamma_{VA} & \gamma_{VV}-\gamma_{00}%
\end{array}
\right), \  \  \
\hat{\gamma}_{(1)}=\left(
\begin{array}
[c]{ccc}%
\gamma_{AA}-\gamma_{00} & \gamma_{AV} & 0\\
\gamma_{VA} & \gamma_{VV}-\gamma_{00} & \frac{56}{3}\gamma_{VG}\\
0 & -\frac{3}{56}\gamma_{GV} & \gamma_{GG}-\gamma_{00}%
\end{array}
\right)  .
\end{equation}
The  values of  $\gamma_{ij}$ are related with the renormalisation of the corresponding operators.  Their values have been obtained from the evolution kernels computed in Ref. \cite{Braun:2009mi}:
\bea
~\gamma_{AA}=\frac{1}{4}C_{F}+\frac{25}{6}C_{A},~~\gamma_{AV}=-\frac{3}{4}C_{F}+\frac{1}{2}C_{A},
\\
~\gamma_{VA}=-\frac{7}{4}C_{F}+\frac{7}{6}C_{A},~~\gamma_{VV}=\frac{31}{12}C_{F}+\frac{13}{6}C_{A}, ~~ \gamma_{VG}=-\frac{n_{f}}{3},
\\
\gamma_{GV}=\frac{8}{3}\left(  8C_{F}-3C_{A}\right)  ,~~\ \gamma_{GG}\mathcal{=}3C_{A}+\frac{2}{3}n_{f}.
\eea
We have checked that Eq.(\ref{Lg8rho}) describing  the evolution of the $\rho$-meson DAs  is in agreement with the expression given in Ref.\cite{Ball:1998ff}.

\section{ Colour-octet corrections in the Coulomb limit}
\label{oct}

In this section we consider  various colour-octet   contributions which
potentially can provide a sizable correction to the amplitudes
derived in the previous sections. The relevance  of the colour-octet
mechanism in decays of $P$-wave charmonia has been discussed in  various
papers, see e.g. Refs.\cite{Bolz:1996wh, Bolz:1997ez, Beneke:2008pi}.  
 In this appendix  we only focus   on  decays $\chi _{c1}\rightarrow V\gamma $ since the corresponding branching rates are
already known from  experiments.

We start our discussion from the longitudinal amplitude $A_{1V}^{\Vert }$
which is suppressed by $\alpha _{s}^{2}$.  In this case the large
correction can be generated by the ultrasoft region where  one of the
gluons in the diagram in Fig.\ref{aiivdiagrams} is ultrasoft.  Although such
contribution might be suppressed by a small factor $v^{2}$ the corresponding
hard kernel is of order $\alpha _{s}$.  In order to simplify our
discussion we consider  the Coulomb limit when $mv^{2}\gg $\ $\Lambda $. In
this case the scales $v$ and $\lambda $ are well separated  $v^{2}\gg
\lambda ^{2}$ and the charmonium state can be considered as a perturbative
Coulomb state.

The simplest way to obtain the subleading amplitude $\left[ A_{1V}^{\Vert }\right] _{oct}\sim \alpha _{s}(\mu )v^{6}$ is just to
consider the ultrasoft limit $k\sim mv^{2}$ in Eq.(\ref{Dh:def}) and to
expand the integrand with respect to small momentum $k$.  In the effective
theory such  a limit  can be associated with the integration over soft gluons
with momenta $k_{s}\sim mv.$ This reduces the description of the heavy quark
sector to potential NRQCD (PNRQCD)\cite{Pineda:1997bj, Pineda:1997ie, Beneke:1997zp, Brambilla:1999qa, Brambilla:1999xf}.  
Schematically,  the factorisation can be described in
the following way. First we integrate over hard modes and obtain  the
effective theory which consists of two sectors: nonrelativistic associated
with the heavy quarks and collinear associated with collinear light
particles. The effective Lagrangian is given by the sum of the NRQCD and
SCET-I($v$) Lagrangians. The latter  describes the
interactions of the ultrasoft $k_{us}\sim m(v^{2},v^{2},v^{2})$ and
hard-collinear particles with momenta $p_{hc}=(\bar{n}\cdot p_{hc},n\cdot
p_{hc},~p_{\perp hc})\sim m(1,v^{2},v).$ At the next step we integrate out
the soft  gluon modes ( $k_{s}^{\mu }\sim mv$) in NRQCD reducing it to
PNRQCD .  To proceed further we integrate over the hard-collinear
particles $p_{hc}\sim m(1,v^{2},v)$ and over ultrasoft modes $k_{us}\sim
m(v^{2},v^{2},v^{2})$ which are still perturbative degrees of freedom. After
that we reduce our description to the effective Lagrangian with $\lambda $%
-collinear  particles with momenta $p_{c}\sim m(1,\lambda
^{4},\lambda ^{2})$.  Only these degrees of freedom can describe an overlap
with the external collinear hadronic states.

\begin{figure}[ptb]
\centering
\includegraphics[width=2.2in]{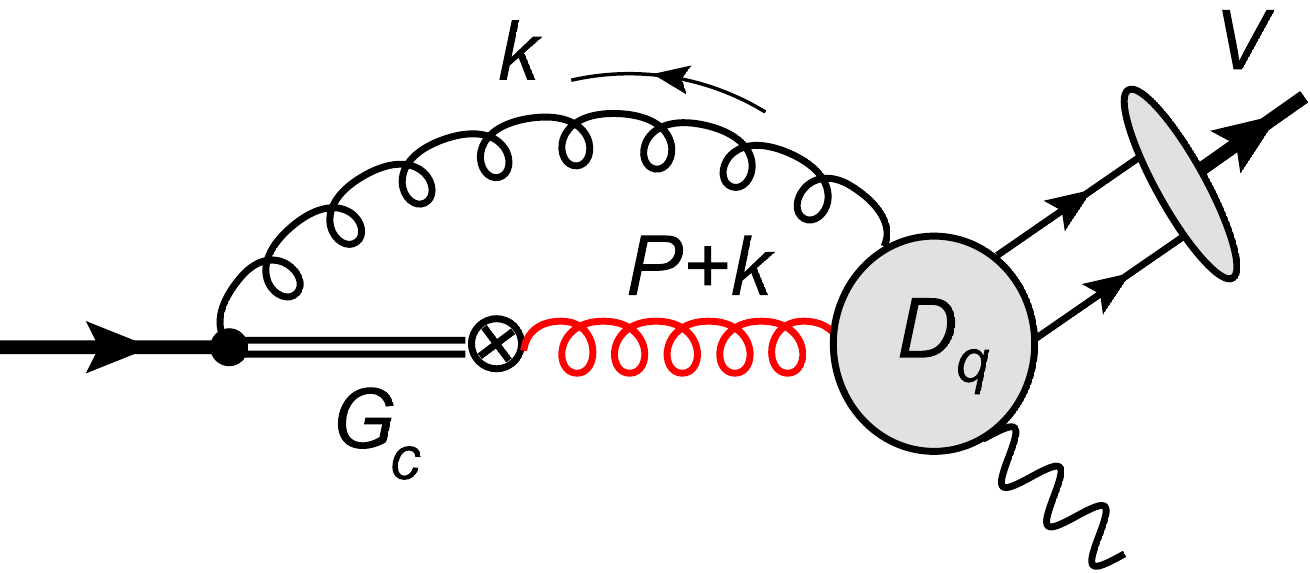}
\caption{The diagram describing the colour-octet
mechanism in the effective theory. $G_{c}$ is the octet propagator in
PNRQCD. The hard gluon is shown by a red curly line. The crossed circle
denotes the vertex of the octet operator (\protect\ref{O8-3S1}). }
\label{col-oct-d1}
\end{figure}

In the following we are not going to build a detailed approach for the
description of the colour-octet contributions. Our task is rather modest: we
would only like  to demonstrate the existence of the colour-octet
contributions with a certain power behaviour. For that purpose we do not need
to perform all required matching calculations in the framework of effective
theory. Instead, we are going to proceed as follows. We assume that the
annihilation of heavy quarks to a hard gluon is described by  simple
hard vertex graph and this gives the NRQCD colour-octet operator 
\begin{equation}
\mathcal{O}^{8}(^{3}S_{1})=\chi _{\omega }^{\dag }\gamma _{\top }^{\beta
}T^{a}\psi _{\omega }.  \label{O8-3S1}
\end{equation}%
The further calculations in the heavy quark sector will be carried out
within the PNRQCD framework.  The subprocesses associated with hard and
hard-collinear scattering will be computed just by expansion with respect
to the small ultrasoft momentum of the corresponding subgraph in QCD. In
other words in these sectors we apply the technique known as expansion by
regions \cite{Beneke:1997zp, Smirnov:2002pj}. This allows us to avoid a detailed discussion of
the SCET formalism and the corresponding matching calculations.  The
structure of the resulting diagram is shown in Fig.\ref{col-oct-d1}. 
The corresponding analytical  expression reads
\bea
 i\varepsilon \lbrack \epsilon _{\chi },\epsilon _{\gamma }^{\ast
},p,q]~(\epsilon _{V}^{\ast }\cdot \omega )\frac{m_{V}}{M_{1}^{2}}\left[
A_{1V}^{\Vert }\right] _{oct}=2(\epsilon _{V}^{\ast }\cdot \omega )Q_{V}%
\sqrt{2}~f_{V}~m_{V}\int_{0}^{1}dx~\phi _{\Vert }^{V}(x)
\nonumber \\ 
\times \int dk~\frac{\left( -i\right) ^{2}\text{Tr}[\mathcal{P}%
_{V}~D_{q}^{\alpha \beta }]}{\left[ k^{2}\right] \left[ 4m^{2}\right] }%
\sqrt{\frac{M}{N_{c}}}\int \frac{d^{3}\boldsymbol{\Delta }}{(2\pi )^{3}}%
\sqrt{\frac{3}{4\pi }}\tilde{R}_{21}(|\boldsymbol{\Delta }|)\frac{1}{4}\text{Tr}\left[ 
\mathcal{P}_{1}(1-\omega )D^{Q}_{\alpha \beta }(E,\Delta _{\top
},k)(1+\omega )\right],  
\label{AII8:def}
\eea
where  $D_{q}^{\alpha \beta }$ denotes the subgraph with hard and
hard-collinear interactions between light quarks and photons, the indices $\alpha $ and $\beta $ are
associated with the gluons ( ultrasoft and hard). In full QCD it is
described by the same expression as in Eq.(\ref{Dq:def}). Now we
assume that  the gluon momentum $k$ is ultrasoft and $D_{q}$ must be
expanded in $k$. The second trace with subgraph $D^{Q}$ describes the heavy
quark sector in PNRQCD and  $R_{21}(|\Delta |)$ is the Coulomb radial wave
function in momentum space, see e.g.  \cite{Beneke:2008pi},
\begin{equation}
\tilde{R}_{21}(|\boldsymbol\Delta |)=-iR_{21}^{\prime }(0)\frac{1024\pi \gamma
_{B}|\boldsymbol{\Delta } |}{\left( 4\boldsymbol{\Delta } ^{2}+\gamma _{B}^{2}\right) ^{3}},~\ \gamma
_{B}=\frac{1}{2}m_{c}\alpha _{s}C_{F}.
\end{equation}%
Many  technical details  have  already been discussed in the literature (see e.g.  Refs.\cite{Brambilla:2004jw, Beneke:2008pi, Brambilla:2012be});   therefore we do not  describe them here.
 The $P$-wave projector  reads%
\begin{equation}
\mathcal{P}_{1}=\frac{1}{2\sqrt{2}}\frac{\Delta _{\top }^{\sigma }}{|\Delta_{\top } |
}\left[ \gamma _{\sigma },\NEG{\epsilon}_{\chi }\right] \gamma _{5}.
\end{equation}%
The analytical expression for the $D_{\alpha\beta}^{Q}$ in Eq.(\ref{AII8:def}) is given by 
\begin{equation}
D_{\alpha\beta}^{Q}(E,\Delta _{\top },k)=\int \frac{d^{3}\mathbf{\Delta }^{\prime }}{(2\pi
)^{3}}~\left[ V_{\mathcal{O}}^{\beta ,b}\right] ~iG_{c}^{(8)}\left[ \Delta
_{\top }^{\prime },\Delta _{\top }+k_{\top },E+(k\omega )\right] ~\left[
V_{g}^{\alpha ,a}\right] ,
\end{equation}%
where $V_{\mathcal{O}}^{\rho ,a}$ and $V_{g}^{\alpha ,b}$denote the operator
and ultrasoft gluon vertices, respectively.  For the colour-octet Coulomb
Green function $G_{c}^{(8)}$ we use the simplified expression suggested in Ref.\cite{Beneke:2008pi} 
\begin{equation}
iG_{c}^{(8)}\left[ \Delta _{\top }^{\prime },\Delta _{\top }+k_{\top
},E+(k\omega )\right] \simeq \frac{(-i)}{E+(k\omega )+\Delta _{\top
}^{2}/m+i\varepsilon }(2\pi )^{3}\delta (\Delta _{\top }^{\prime }-\Delta
_{\top }).
\end{equation}%
 The  ultrasoft gluon vertex $V_{g}$ is generated by the chromoelectric dipole
interaction in the effective PNRQCD Lagrangian  
\begin{equation}
\mathcal{L}_{int}=g\int d^{3}\mathbf{x}~S_{\omega}^{\dag }(\mathbf{x})\left[ ~\vec{x}%
\cdot \vec{E}(t,0)\right] ~O_{\omega}(\mathbf{x}),
\end{equation}%
where $S_{\omega}$ and $O_{\omega}$ denote the quark-antiquark singlet and octet fields,
respectively, see e.g. \cite{Brambilla:2004jw}. In momentum space this gives 
\begin{eqnarray}
D^{Q}_{\alpha \beta }(E,\Delta _{\top },k) &=&\left[ igT^{b}\gamma _{\top\beta
}\right] _{V_{\mathcal{O}}}~\left[ gT^{a}\left\{ \omega_\alpha k_{\top \lambda}-g_{\top \alpha \lambda }(\omega k)\right\} \frac{%
\partial }{\partial \Delta _{\top }^{\lambda }}\right] _{V_{g}}\frac{(-i)}{%
\left[ E+(k\omega )+\Delta _{\top }^{2}/m+i\varepsilon \right] }  \notag \\
&=&\left[ igT^{b}\gamma _{\top \beta }\right] _{V_{\mathcal{O}%
}}~gT^{a}\left\{ \omega _{\alpha }k_{\top \lambda}-g_{\top \alpha
\lambda }(\omega k)\right\} \frac{(-2)}{m}\frac{i\Delta _{\top }^{\lambda }}{%
\left[ E+(k\omega )+\Delta _{\top }^{2}/m+i\varepsilon \right] ^{2}}.
\end{eqnarray}
 The trace in Eq.(\ref{AII8:def}) yields%
\begin{eqnarray}
\frac{1}{4}\text{Tr}\left[ \mathcal{P}_{1}(1-\omega )D^{Q}_{\alpha\beta}(E,\Delta _{\top
})(1+\omega )\right] =\frac{\delta ^{ab}}{2}\frac{g\left\{ \omega _{\alpha
}k_{\top\lambda }-g_{\top \alpha \lambda }(\omega k)\right\} }{\left[
E+(k\omega )+\Delta _{\top }^{2}/m+i\varepsilon \right] ^{2}}  \frac{2}{m}\frac{1}{\sqrt{2}}\frac{\Delta _{\top }^{\lambda }\Delta
_{\top }^{\sigma }}{|\Delta_\top |}i\varepsilon \lbrack \omega \epsilon _{\chi
}\sigma \beta ],
\end{eqnarray}
and  we obtain
\bea
~\int \frac{d^{3}\boldsymbol{\Delta }}{(2\pi )^{3}}\tilde{R}_{21}(|\mathbf{\Delta } |)%
\frac{1}{4}\text{Tr}\left[ \mathcal{P}_{1}(1-\omega )D^{Q}_{\alpha\beta}(E,\Delta _{\top},k)
(1+\omega )\right]
\nonumber \\
=-\frac{\delta ^{ab}}{\sqrt{2}}\ \frac{g}{3m}\int \frac{d^{3}\boldsymbol{%
\Delta }}{(2\pi )^{3}}|\boldsymbol{\Delta }|\tilde{R}_{21}(|\mathbf{\Delta }|)\frac{i\varepsilon \lbrack \omega \epsilon _{\chi }\sigma \beta ]\left\{
\omega _{\alpha }k_{\top }^{\sigma }-(g_{\top })_{\alpha}^{ \sigma }(\omega
k)\right\} }{\left[ E+(k\omega )+\Delta _{\top }^{2}/m+i\varepsilon \right]^{2}},
\eea
where we used rotational invariance  in order
to rewrite \ $\Delta _{\top }^{\lambda }\Delta _{\top }^{\sigma }\rightarrow -|\mathbf{%
\Delta }|^{2}g_{\top }^{\lambda \sigma }/3$.  Substituting this into Eq.(%
\ref{AII8:def}) and computing the colour trace we obtain%
\bea
&&i\varepsilon \lbrack \epsilon _{\chi },\epsilon _{\gamma }^{\ast
},p,q]~(\epsilon _{V}^{\ast }\cdot \omega )\frac{m_{V}}{M_{1}^{2}}\left[
A_{1V}^{\Vert }\right] _{oct}=(\epsilon _{V}^{\ast }\cdot \omega)Q_{V}
\frac{ i f_{V}m_{V}}{m^3}\sqrt{\frac{M}{N_{c}}}\sqrt{\frac{3}{4\pi }}\frac{C_{F}}{6\pi ^{2}}
\alpha _{s}(\mu _{h})\alpha _{s}(\mu _{us})%
\notag \\ 
&&\times  
\int_{0}^{1}dx~\phi _{\Vert }^{V}(x)\int \frac{d^{3}\mathbf{\Delta }}{(2\pi
)^{3}}|\boldsymbol{\Delta }|\tilde{R}_{21}(|\mathbf{\Delta }|)  
%\notag \\&&\times 
\int d^{4}k~\frac{\text{Tr}[\mathcal{P}_{V}~D_{q}^{\alpha \beta }]}{%
\left[ k^{2}\right]  }\frac{i\varepsilon \lbrack
\omega \epsilon _{\chi }\sigma \beta ]g\left\{ \omega _{\alpha }k_{\top
}^{\sigma }-(g_{\top })_\alpha^{ \sigma }(\omega k)\right\} }{\left[ E+(k\omega
)+\Delta _{\top }^{2}/m+i\varepsilon \right] ^{2}},~  \label{aIIV-1}
\eea
where the Lorentz indices $\alpha $ and $\beta $ correspond to the
ultrasoft and hard (i. e. colour-octet operator) gluon respectively. The
expansion of the light quark term Tr$[\mathcal{P}_{V}~D_{q}]$ is carried out
by expanding the  hard-collinear and hard propagators to the
next-to-leading order 
\begin{equation}
\frac{(p_{i}+k)}{\left[ (p_{i}+k)^{2}\right] }\simeq \frac{\NEG{p}_{i}}{%
\left[ 2(kp_{i})\right] }+\frac{\NEG{k}}{\left[ 2(kp_{i})\right] }-\frac{%
k^{2}}{\left[ 2(kp_{i})\right] ^{2}},
\end{equation}%
\begin{equation}
\frac{\NEG{p}_{i}+\NEG{q}+\NEG{k}}{\left[ \left( p_{i}+q+k\right) ^{2}\right]
}\simeq \frac{\left( \NEG{p}_{i}+\NEG{q}\right) }{\left[ 2\left(
p_{i}q\right) \right] }+\frac{\NEG{k}}{\left[ 2\left( p_{i}q\right) \right] }%
-\frac{\left( \NEG{p}_{2}+\NEG{q}\right) }{\left[ 2\left( p_{i}q\right) %
\right] }\frac{2k(p+q)}{\left[ 2\left( p_{i}q\right) \right] },
\label{hard-exp}
\end{equation}%
where $p_{i}=\{xp,\bar{x}p\}$. The result can be written as 
\bea
{\text{Tr}[\mathcal{P}_{V}~D_{q}^{\alpha \beta }]}&\simeq &\frac{1}{\left[ 8m^{3}\right]}\frac{1}{%
\left[ (kp)\right] }\left( p^{\beta }\left\{ p^{\alpha }(\varepsilon _{\gamma
}k)-\varepsilon _{\gamma }^{\alpha }(kp)\right\} \frac{2}{x}\right.
\nonumber \\ &&
+q^{\beta }\left\{ (\varepsilon _{\gamma }k)p^{\alpha }-(kp)\varepsilon
_{\gamma }^{\alpha }\right\} \frac{1-2x}{2x^{2}}+\varepsilon _{\gamma
}^{\beta }\left\{ (kp)q^{\alpha }-p^{\alpha }(kq)\right\} \frac{3-2x}{2x^{2}}
\nonumber \\ &&
\left. +\left[ (pq)\left\{ \varepsilon _{\gamma }^{\alpha }k^{\beta
}-g^{\alpha\beta }(\varepsilon _{\gamma }k)\right\} +p^{\beta }\left\{
q^{\alpha }(\varepsilon _{\gamma }k)-\varepsilon _{\gamma }^{\alpha
}(kq)\right\} \right] \frac{1+2x}{2x^{2}}\right) +\left( x\rightarrow \bar{x}%
\right) .
\eea
Notice that this  expression vanishes if it is contracted with the
ultrasoft gluon momentum $k^{\alpha }$, %
\begin{equation}
{k_{\alpha }\text{Tr}[\mathcal{P}_{V}~D_{q}^{\alpha \beta }]}=0,
\end{equation}
as it is required by the gauge invariance. Substituting this into Eq.(\ref%
{aIIV-1}) and performing contractions of the Lorentz indices we obtain 
\begin{equation}
\left[ A_{1V}^{\Vert }\right] _{oct}\sim \alpha _{s}(\mu _{h})\alpha
_{s}(\mu _{us})~\frac{f_{V}M_{1}^{2}}{m^{6}}\sqrt{M_{1}}\int_{0}^{1}dx~\frac{%
\phi _{\Vert }^{V}(x)}{x}\int \frac{d^{3}\mathbf{\Delta }}{(2\pi )^{3}}~%
\tilde{R}_{21}(|\mathbf{\Delta }|)|\mathbf{\Delta }|~J_{us},  \label{AIIVoct}
\end{equation}%
with%
\begin{equation}
J_{us}=\int dk~\frac{1}{\left[ k^{2}\right] \left[ E+(k\omega )+\Delta
_{\top }^{2}/2m+i\varepsilon \right] ^{2}}\left\{ \frac{3}{2x}(k\bar{n}%
)-2(k\omega )-2(k\bar{n})\right\} .
\end{equation}%
From this result one can easily get the scaling behaviour of the
colour-octet amplitude. The power of $\lambda $ is again provided by  the leading twist operator in Eq.(\ref{scaling}). The
power of velocity $v$ can be obtained from Eq.(\ref{AIIVoct}) using that $\Delta \sim mv$, $k\sim mv^{2}$ and $~\tilde{R}_{21}(|\mathbf{\Delta }|)\sim
v^{0}$. This gives%
\begin{equation}
\left[ A_{1V}^{\Vert }\right] _{oct}\sim \alpha _{s}(\mu _{h})\alpha
_{s}(\mu _{us})~v^{6}\lambda ^{2}.  \label{AIIoct-1}
\end{equation}%

The expression for the amplitude in Eq.(\ref{AIIVoct}) includes the
divergent collinear convolution integral 
\begin{equation}
\int_{0}^{1}dx~\frac{\phi _{\Vert }^{V}(x)}{x^{2}}=\left[ \phi _{\Vert
}^{V}(0)\right] ^{\prime }\int_{0}^{1}~\frac{dx}{x}+\int_{0}^{1}dx~\frac{
\phi _{\Vert }^{V}(x)-x\left[ \phi _{\Vert }^{V}(0)\right] ^{\prime }}{x^{2}},
\end{equation}%
where it was used  that $\ \phi _{\Vert }^{V}(x)-x\left[ \phi _{\Vert }^{V}(0)
\right] ^{\prime }\overset{x\rightarrow 0}{\sim }\mathcal{O}(x^{2})$.  This indicates
that  there must be one more term which can be associated with the endpoint region
where the collinear fraction $x$ is small. Such contribution can be obtained
in SCET-I($v$) if we consider appropriate configuration for $D_{q}$ where
the outgoing quark-antiquark is not collinear as before but consists of 
the $v-$ultrasoft quark ($p_{q}\sim mv^{2}$) and collinear antiquark fields.
 In the Coulomb limit  such a pair corresponds to the configuration when the
collinear fraction of the outgoing quark is small $x\sim v^{2}$ but still
large compared to the nonpertubative QCD scale%
\begin{equation}
x\sim v^{2}\gg \lambda^2 \sim \Lambda /m.
\end{equation}%
The $v-$ultrasoft quark field can be matched onto $\lambda $-collinear field
with the small collinear fraction. Hence the corresponding matrix element
still can be understood as $\lambda -$collinear matrix element in the region
where $\lambda \ll x\ll 1$ and therefore it can be computed in terms of
light-cone DA  which must be expanded with respect to $x$ in order to avoid
large power of scale $v$.  A similar situation has been  considered
in Ref.\cite{Beneke:2008pi}.  In the present case the computation of the corresponding
colour-octet contribution is very similar to the one carried out above, only 
the expansion of the light quark part $D_{q}$ is  different. Now one has
to take into account that 
\begin{equation}
xp\sim k_{us}\sim v^{2}.
\end{equation}%
This allows one to expand the meson DA: 
\begin{equation}
\phi _{\Vert }^{V}(x)=x\left[ \phi _{\Vert }^{V}(0)\right] ^{\prime }+%
\mathcal{O}(v^{4}).
\end{equation}%
The amplitude now reads%
\begin{equation*}
i\varepsilon \lbrack \epsilon _{\chi },\epsilon _{\gamma }^{\ast
},p,q]~(\epsilon _{V}^{\ast }\cdot \omega )\frac{m_{V}}{M_{1}^{2}}\left[
A_{1V}^{\Vert }\right] _{oct,~endp}\sim (\epsilon _{V}^{\ast }\cdot \omega
)~f_{V}m_{V}~\frac{1}{m^{3}}\sqrt{M}
\end{equation*}%
\begin{equation}
~\times \int \frac{d^{3}\mathbf{\Delta }}{(2\pi )^{3}}\tilde{R}_{21}(|%
\mathbf{\Delta }|)|\mathbf{\Delta }|~\int_{0}^{\eta }dx~x~\left[ \phi
_{\Vert }^{V}(0)\right] ^{\prime }\int dk~\text{Tr}[\mathcal{P}_{V}D^{\alpha\beta}_{q}]%
\frac{i\varepsilon \lbrack \omega \epsilon _{\chi }\lambda \beta ]\left\{
\omega _{\alpha }k_{\top }^{\lambda }-(g_{\top })_\alpha^{\lambda }(\omega
k)\right\} }{\left[ k^{2}\right] \left[ E+(k\omega )+\Delta _{\top
}^{2}/m+i\varepsilon \right] ^{2}},  \label{AIIVoctendp}
\end{equation}%
where we introduce explicitly the regularisation cutoff~$\eta $ which
restricts  the collinear fraction $x$ to be small. A similar
regularisation must also be implied for the  IR-divergent integral in Eq.(%
\ref{AIIVoct}). The expanded light quark trace reads%
\begin{equation}
\text{Tr}[\mathcal{P}_{V}D_{q}^{\alpha \beta }]\simeq \frac{\text{Tr}\left[ 
\mathcal{P}_{V}\gamma ^{\alpha }(xp+k)\epsilon _{\gamma }^{\ast }q\gamma
^{\beta }\right] }{\left[ \left( xp+k\right) ^{2}\right] \left[ 2q(xp+k)%
\right] }+\frac{\text{Tr}\left[ \mathcal{P}_{V}\epsilon _{\gamma }^{\ast
}q\gamma ^{\alpha }q\gamma ^{\beta }\right] }{\left[ 2x(pq)\right] \left[
2q(xp+k)\right] }+\frac{\text{Tr}\left[ \mathcal{P}_{V}\epsilon _{\gamma
}^{\ast }q\gamma ^{\beta }(-p)\gamma ^{\alpha }\right] }{\left[ 2x(pq)\right]
\left[ 2(kp)\right] }.
\end{equation}%
This gives 
\begin{equation}
\text{Tr}[\mathcal{P}_{V}D_{q}^{\alpha \beta }]\sim v^{-4}.
\end{equation}%
Hence taking into account that $\ dx~x\sim v^{4}$ we get%
\begin{equation}
\left[ A_{1V}^{\Vert }\right] _{oct,~endp}\sim \alpha _{s}(\mu _{h})\alpha
_{s}(\mu _{us})v^{6}\lambda ^{2},  \label{AIIVoct-2}
\end{equation}%
i.e. the endpoint amplitude is of the same order as the colour-octet term in Eq.(\ref{AIIVoct}). We expect that the regularisation scale $\eta $ cancels in
the sum of the contributions in Eqs.(\ref{AIIVoct}) and (\ref{AIIVoctendp}).
 The check of this statement can be done by explicit calculation  but such consideration is beyond the scope of the present work.

\begin{figure}[ptb]
\centering
\includegraphics[width=5.6936in]
{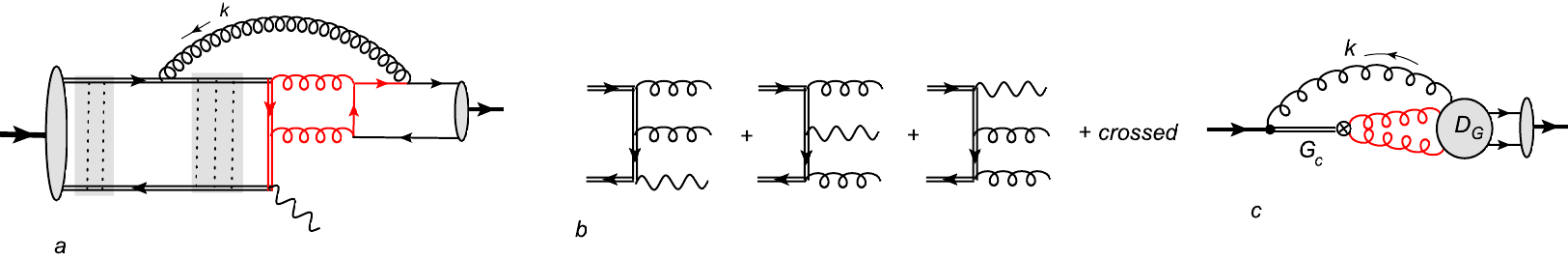}
\caption{The diagrams
describing the colour-octet mechanism in the vector meson production. $(a)$ An
example of the diagram  with gluons in the intermediate state. The red
lines show the hard and hard-collinear particles in SCET-I($v$). The dotted
lines in the grey boxes denote Coulomb gluon exchanges. $(b)$ The diagrams
describing the hard matching.  $(c)$ The diagram in the effective theory
describing the matrix element $G_{V}^{\Vert }$ in Eq.(\ref{GVert}) .  }
\label{col-oct-d2}
\end{figure}

\begin{figure}[ptb]
\centering
\includegraphics[width=5.227in]
{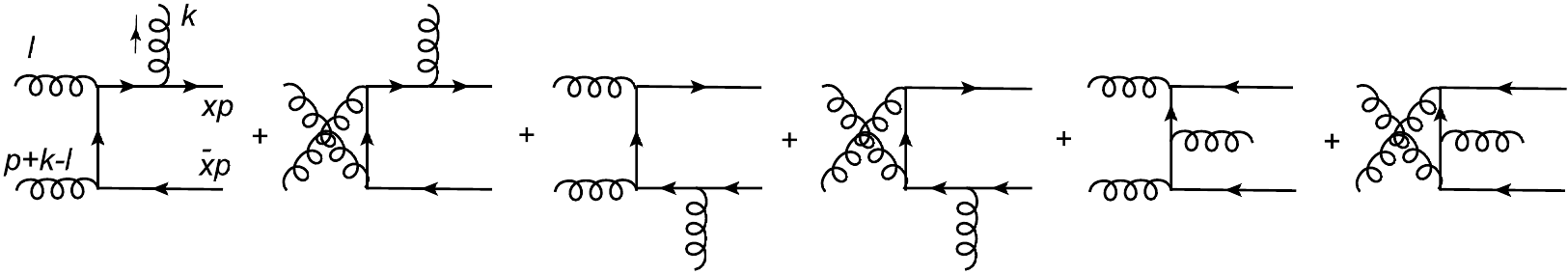}
\caption{The
light quark subdiagrams describing the function $\Sigma _{q}$ in Eq.(\ref{DG:def}). }
\label{sigmaq}
\end{figure}

Even if the the colour-octet mechanism described above is sufficiently large it cannot explain the  large ratio 
$Br[\chi_{1c}\rightarrow \omega_{\Vert}\gamma]/Br[\chi_{1c}\rightarrow \rho_{\Vert}\gamma]$. 
It is natural to expect  that for an isosinglet meson  there could be  a specific,  
relatively large  contribution  which can  explain  the  enhancement of the ratio.  
Such mechanism  can be related with the contribution  which involves the heavy
quark term of the electromagnetic  current in the definition of the
amplitude Eq.(\ref{Ame}).  In this case  the light meson state 
is produced through  the  interaction of collinear and soft gluons.  In order to have minimal power of $\lambda$ the gluons couple to collinear quarks.  
The corresponding diagrams are shown in Fig.\ref{col-oct-d2}$(a)$. 
The hard subprocess describes the
annihilation of a heavy quark-antiquark pair into a photon and two
hard-collinear gluons in an octet state; see Fig.\ref{col-oct-d2}$(b)$. Performing the hard
factorisation one obtains%
\begin{equation}
\left\langle V(p)\right\vert \epsilon _{\gamma }^{\ast }\cdot
J_{Q}(0)\left\vert \chi _{c1}(P)\right\rangle =i\varepsilon \lbrack \epsilon
_{\chi },\epsilon _{\gamma }^{\ast },p,q]~(\epsilon _{V}^{\ast }\cdot \omega
)\frac{m_{V}}{M_{1}^{2}}~iA_{Q}^{\Vert },
\end{equation}%
\begin{equation}
~iA_{Q}^{\Vert }~=ee_{Q}~\frac{\pi \alpha _{s}(\mu _{h})}{2m^{4}}%
\int_{0}^{1}d\tau ~\frac{1}{\tau \bar{\tau}}G_{V}^{\Vert }(\tau ),
\end{equation}%
where $\tau $ is the collinear fraction of the gluon momenta and the function $G_{V}^{\Vert }(\tau )$ is defined as  
the matrix element of the following colour-octet operator
\begin{align}
& i\varepsilon \lbrack \epsilon _{\chi },\epsilon _{\gamma }^{\ast
},p,q]~(\epsilon _{V}^{\ast }\cdot \omega )\frac{m_{V}}{M_{1}^{2}}G_{\Vert
}^{V}(\tau )=\notag  \\
& p_{-}\int \frac{d\lambda }{\pi }e^{-i\lambda p_{-}(2\tau -1)}~\epsilon
_{\gamma }^{\sigma \ast }~\left\langle V(p)\right\vert T\left\{
O_{8}^{\sigma }(^{3}S_{1})d^{abc}G_{\bar{n} }^{a\mu}(\lambda \bar{n})G_{\bar{%
n}\mu }^{b}(-\lambda \bar{n})\right\} \left\vert \chi _{c1}(P)\right\rangle ,
\label{GVert}
\end{align}%
where $p_-\equiv (p\bar n)$ and  for simplicity, we do not show the Wilson lines in the hard-collinear
gluon operator ( $G_{\mu\nu}$ is the gluon strength tensor), using the notation of Eq.(\ref{Gnbar}). In the Coulomb limit this matrix element can be computed in
the similar way as it was discussed above; see Fig.\ref{col-oct-d2}$(c)$.  We
will not repeat the details and provide the complete analytical expression
(up to irrelevant constants) 
\begin{equation}
\left\langle V(p)\right\vert \epsilon _{\gamma }^{\ast }\cdot
J_{Q}(0)\left\vert P\right\rangle \sim \left( \epsilon _{V}^{\ast }\cdot 
\bar{n}\right) ~ee_{Q}\sqrt{M}\frac{~f_{V}m_{V}}{m^{5}}\alpha _{s}(\mu
_{h})\alpha _{s}(\mu _{hc})\alpha _{s}(\mu _{us})  \label{AQ-1}
\end{equation}%
\begin{equation}
\times \int dx\phi _{\Vert }^{V}(x)\int_{0}^{1}\frac{d\tau }{\tau \bar{\tau}}%
\int \frac{d^{3}\mathbf{\Delta }}{(2\pi )^{3}}~R_{21}(|\mathbf{\Delta }|)||%
\mathbf{\Delta }|\int \frac{d^{4}k}{(2\pi )^{4}}D_{G}^{\alpha }(k)\frac{%
i\varepsilon \lbrack \omega \epsilon _{\chi }\lambda \epsilon _{\gamma
}^{\ast }]\left\{ \omega _{\alpha }k_{\top }^{\lambda }-(g_{\top })_\alpha^{
\lambda }(\omega k)\right\} }{\left[ k^{2}\right] \left[ E+(k\omega )+\Delta
_{\top }^{2}/m+i\varepsilon \right] ^{2}},
\end{equation}%
where the subdiagram $D_{G}$ describes the gluon loop and interaction of the
ultrasoft gluon with collinear quarks. The corresponding expression reads 
\begin{equation}
D_{G}^{\alpha }(k)=\int d^{D}l\frac{V_{gg}^{\beta \sigma }(l)\text{Tr}\left[ 
\mathcal{P}_{V}\Sigma _{q}^{\alpha \beta \sigma }\right] }{\left[ l^{2}%
\right] [(p+k-l)^{2}]},  \label{DG:def}
\end{equation}%
with the gluon operator vertex 
\begin{equation}
V_{gg}^{\beta \sigma }\simeq \left\{ l_{-}(p_{-}-l_{-})g_{\bot }^{\beta
\sigma }-(p_{-}-l_{-})l_{\bot }^{\sigma }\bar{n}^{\beta }+l_{-}l_{\bot
}^{\beta }\bar{n}^{\sigma }+l_{\bot }^{2}\bar{n}^{\beta }\bar{n}^{\sigma
}\right\} \delta (l_{-}/p_{-}-\tau ).
\end{equation}%
In the expression for the light quark part Tr$\left[ \mathcal{P}_{V}\Sigma
_{q}\right] $ we are projecting the outgoing quark-antiquark pair to the twist-2
collinear operator  which yields the  meson DA $\phi _{\Vert }^{V}(x)$,  and we expand the resulting QCD expression
with respect to small $v$-ultrasoft momentum $k$. For the hard-collinear
gluon loop momentum $l$ we assume 
\begin{equation}
l_{-}\sim v^{0},~\ l_{+}\sim v^{2},~l_{\bot }\sim v.
\end{equation}%
The analytical expression for $\Sigma _{q}$ can be obtained from the
corresponding QCD diagrams by expansion with respect to ultrasoft momentum $%
k $, see Fig.\ref{sigmaq}. 
%
%\begin{eqnarray}
%\left. \Sigma _{q}^{\alpha \beta \sigma }\right\vert _{k\sim mv^{2}} &=&%
%\frac{\gamma ^{\alpha }(k+xp)\gamma ^{\beta }(xp+k-l)\gamma ^{\sigma }}{%
%\left[ (k+xp)^{2}\right] \left[ (k+xp-l)^{2}\right] }+\frac{\gamma ^{\alpha
%}(k+xp)\gamma ^{\sigma }(l-\bar{x}p)\gamma ^{\beta }}{\left[ (k+xp)^{2}%
%\right] \left[ (l-\bar{x}p)^{2}\right] }  \notag \\
%&&+\frac{\gamma ^{\beta }(xp-l)\gamma ^{\sigma }(-k-\bar{x}p)\gamma ^{\alpha
%}}{\left[ (xp-l)^{2}\right] \left[ (k+\bar{x}p)^{2}\right] }+\frac{\gamma
%^{\sigma }(l-k-\bar{x}p)\gamma ^{\beta }(-k-\bar{x}p)\gamma ^{\sigma }}{%
%\left[ (l-k-\bar{x}p)^{2}\right] \left[ (k+\bar{x}p)^{2}\right] }  \notag \\
%&&+\frac{\gamma ^{\beta }(xp-l)\gamma ^{\alpha }(xp+k-l)\gamma ^{\sigma }}{%
%\left[ (xp-l)^{2}\right] \left[ (xp+k-l)^{2}\right] }+\frac{\gamma ^{\sigma
%}(l-k-\bar{x}p)\gamma ^{\alpha }(l-\bar{x}p)\gamma ^{\beta }}{\left[ (l-k-%
%\bar{x}p)^{2}\right] \left[ (l-\bar{x}p)^{2}\right] }.
%\end{eqnarray}%
One obtains that the leading-order term scales
as 
\begin{equation}
V_{gg}^{\beta \sigma }(l)\text{Tr}\left[ \mathcal{P}_{V}\Sigma _{q}^{\alpha
\beta \sigma }\right] \sim v^{0}.  \label{VtrS}
\end{equation}%
Performing  an expansion with respect to $k$  one finds  that individual
diagrams have the terms of order $v^{-2}$ which appear from the graphs where
the ultrasoft photon is attached to the external quark lines. However such
contributions cancel because of colour neutrality of the outgoing
quark-antiquark pair. This means that the interaction of the ultrasoft gluon with hard-collinear
quarks is described by the subleading interactions in the SCET-I($v$)
Lagrangian. The analytical result for to the trace in Eq.(\ref{VtrS}) is
somewhat lengthy and we will not write it here. Using the estimate 
(\ref{VtrS}) one finds that
\begin{equation}
D_{G}(k)\sim v^{0},
\end{equation}%
and using this in Eq.(\ref{AQ-1}) one obtains 
\begin{equation}
\left\langle V(p)\right\vert \epsilon _{\gamma }^{\ast }\cdot
J_{Q}(0)\left\vert P\right\rangle \sim A_{Q}^{\Vert }\sim \alpha _{s}(\mu
_{h})\alpha _{s}(\mu _{hc})\alpha _{s}(\mu _{us})~\lambda ^{2}v^{6}\text{.}
\end{equation}%
This estimate is only suppressed by the hard-collinear coupling $\alpha
_{s}(\mu _{hc})$ compared to the estimate of the colour-octet amplitudes in
Eqs. (\ref{AIIoct-1}) and (\ref{AIIVoct-2}). Taking into account that the
hard-collinear scale $\mu _{hc}\sim mv$  we obtain $\alpha _{s}(\mu
_{hc})\sim v$ . Therefore 
\begin{equation}
A_{Q}^{\Vert }\sim \alpha _{s}(\mu _{h})\alpha _{s}(\mu _{us})~\lambda ^{2}v^{7}.
\end{equation}%

We next discuss the colour-octet contribution in the amplitude with a transverse meson.  In
this case we consider the diagram as in Fig.\ref{col-oct-d1}  projecting the
two-quark state to the twist-3 DA.  In the general case one has to consider two- and three-point matrix elements which depend on  
all possible DAs.  For simplicity we only consider the two-point matrix elements  and pick up the
so-called Wandzura-Wilzcek contribution which is completely determined by
the leading twist DA $\phi _{\Vert }^{V}$. 
The description of the corresponding matrix element and corresponding twist-3 DAs  can be found  in Appendix A.  The heavy quark part in this case is the
same as in Eq.(\ref{aIIV-1}); therefore we can write 
\bea
~i\varepsilon \lbrack \epsilon _{V}^{\ast },\epsilon _{\gamma }^{\ast
},p,q]~(\epsilon _{\chi }\cdot q)\frac{1}{M^2_1}\left[ A_{V}^{\bot }\right]
_{oct}\sim ~\frac{f_{V}~m_{V}}{m^{3}}\alpha _{s}(\mu _{h})\alpha
_{s}(\mu _{us})\sqrt{M}
\nonumber \\
\times \int \frac{d^{3}\mathbf{\Delta}}{(2\pi )^{3}}|\boldsymbol{\Delta }|
\tilde{R}_{21}(|\mathbf{\Delta }|)\int_{0}^{1}dx~\int d^{4}k~\frac{\text{Tr}[
\mathcal{P}_{V}^{tw3}D^{\alpha\beta}_{q}]}{\left[ k^{2}\right] }
\frac{i\varepsilon \lbrack \omega \epsilon _{\chi }\sigma \beta ]g\left\{
\omega _{\alpha }k_{\top }^{\sigma }-(g_{\top })_\alpha^{ \sigma }(\omega
k)\right\} }{\left[ E+(k\omega )+\Delta _{\top }^{2}/m+i\varepsilon \right]
^{2}}, 
\label{AVperp-0}
\eea
where the integral over $dx$ denotes the integral over the collinear
fraction of the outgoing quarks.   The twist-3 projector reads 
\bea
\text{Tr}[\mathcal{P}_{V}^{tw3}D^{\alpha\beta}_{q}]=
\left\{  g_V^{v}(x)(\varepsilon_V^*)_{\sigma\bot}
+\int_0^x du\left[ g_V^{v}(u)- \phi_V^\Vert(u)\right]  
p_{\sigma}(\varepsilon_V^*)_{\bot}^{\rho}\frac{\partial}{\partial p_{\bot}^{\rho}
}\right\}  \frac{1}{4}\text{Tr}\left[  \gamma^{\sigma}D^{\alpha\beta}_q\right]
\nonumber \\
-\frac14 g^{a\, \prime}_V(x)\frac18 \text{Tr}\left[ \gamma_5\gamma^\sigma\Dsl{\varepsilon}_V^*\ns\nbs\right] 
\left. \frac{1}{4}\text{Tr}\left[  \gamma^{\sigma}\gamma_5D^{\alpha\beta}_q\right]  \right |_{p_{\perp}=0},
\eea
where $g^{a\, \prime}_V(x)=dg^{a}_V(x)/dx $.
The expression for  $D_q$ is given in Eq.(\ref{Dq:def}) but  external quark momenta $p_{1,2}$ now  have transverse components
\begin{equation}
p_{1}=x_{1}p+p_{\bot},~p_{2}=x_{2}p-p_{\bot}.
\end{equation}
The explicit expressions for twist-3 DAs $g^v_V$ and $g_V^a$ are given in Eqs.(\ref{gv}) and (\ref{ga}). 
Computing the trace and expanding with respect to small $k$ one obtains
\begin{equation}
\frac{1}{4}\text{Tr}[\mathcal{P}_{V}^{tw3}D^{\alpha\beta}_{q}]\simeq\frac{1}{4}
\frac{\text{Tr}\left[\not \epsilon _{V}^{\ast}\gamma_{\bot}^{\alpha}\gamma_{\bot}^{\beta}
\not \epsilon _{\gamma}^{\ast}\right]  }{\left[  -2(pk)\right]  }
\left\{  \frac{1}{2\bar{x}^{2}}%
\int_{x}^{1}dv\frac{\phi(v)}{v}-\frac{1}{\bar{x}}\frac{1}{4}g_V^{a\, \prime}(x)+(x\rightarrow \bar x)\right\} .
\label{Tr-tw3}
\end{equation}
The dominant contribution arises from the diagrams with soft gluon emission from external quark lines.  For the twist-3 case  such contributions do not cancel 
because of the  off-shellness  of the external particles.   Substituting (\ref{Tr-tw3}) into Eq.(\ref{AVperp-0}) gives
\bea
\left[ A_{1V}^{\bot }\right]_{oct}\sim \frac{f_{V}m_V M^2_{1}}{m^{7}}\alpha _{s}(\mu _{h})\alpha
_{s}(\mu _{us})\sqrt{M_1} \int_{0}^{1}dx~\left\{  \frac{1}{2\bar{x}^{2}}%
\int_{x}^{1}dv\frac{\phi(v)}{v}-\frac{1}{\bar{x}}\frac{1}{4}g_{V}^{a\, \prime}(x)+(x\rightarrow \bar x)\right\}
\nonumber \\
\times \int \frac{d^{3}\mathbf{\Delta}}{(2\pi )^{3}}|\boldsymbol{\Delta }|
\tilde{R}_{21}(|\mathbf{\Delta }|)\int d^{4}k~
\frac{ (\omega k)}{\left[ k^{2}\right] [-(kn)] \left[ E+(k\omega )+\Delta _{\top }^{2}/m+i\varepsilon \right]
^{2}}.
\label{A1Vperp}
\eea
From this expression one finds the following  estimate 
\bea
\left[ A_{1V}^{\bot }\right]_{oct}\sim \alpha _{s}(\mu _{h})\alpha_{s}(\mu _{us})\lambda^4 v^4.
\eea
Let us also note that the collinear  integrals  in Eq.(\ref{A1Vperp})  have the endpoint singularities that can be easily seen using an explicit expression for the functions
$g^{a,v}_V$.  Therefore in this case one  also has  to include  endpoint contributions which  will not be considered here.

\end{appendix}

\end{document}